\documentstyle[12pt,epsf,amssymb]{article}

\renewcommand{\theequation}{\arabic{section}.\arabic{equation}}
\renewcommand{\thefootnote}{\fnsymbol{footnote}}

\makeatletter
\def\slash#1{{\mathpalette\c@ncel{#1}}} 
\makeatother

\begin{document}

\begin{titlepage}
\begin{flushright}
\begin{tabular}{l}
CERN--TH/98--162\\
NORDITA--98--41--HE\\
hep-ph/9805422
\end{tabular}
\end{flushright}
\vskip0.5cm
\begin{center}
  {\Large \bf
        Exclusive Semileptonic and Rare B-Meson Decays\\[5pt] in QCD
  \\}
\vspace{1cm}
{\sc Patricia~Ball}${}^{1,}$\footnote{Heisenberg Fellow}, 
{\sc V.M.~Braun}${}^{2,}$\footnote{
On leave of absence from
St.Petersburg Nuclear Physics Institute, 188350 Gatchina,
Russia}
\\[0.3cm]
\vspace*{0.1cm} ${}^1$ {\it CERN--TH, CH--1211 Gen\`eve 23, Switzerland
} \\[0.3cm]
\vspace*{0.1cm} ${}^2${\it NORDITA, Blegdamsvej 17, DK--2100 Copenhagen,
Denmark}
\\[1.3cm]



\vfill
 
 {\large\bf Abstract:\\[10pt]} \parbox[t]{\textwidth}{
We present the first complete results for the semileptonic and rare radiative 
form factors of $B$-mesons weak decay into a light vector-meson
($\rho,\omega,K^*,\phi$) in the light-cone sum-rule approach. 
The calculation includes radiative corrections, 
higher-twist corrections and SU(3)-breaking effects. The theoretical 
uncertainty is investigated in detail. A simple parametrization of the 
form factors is given in terms of three parameters each. 
We find that the form factors observe several relations inspired by 
heavy-quark symmetry.
}
  \vskip1cm
{\em Submitted to Physical Review D}
\end{center}
\end{titlepage}

\renewcommand{\thefootnote}{\arabic{footnote}}
\setcounter{footnote}{0}

\section{Introduction}
\setcounter{equation}{0}

The  challenge to understand the physics of CP-violation related to 
the structure of the CKM mixing matrix in (and beyond) the Standard
Model is fuelling  an impressive experimental programme for the study 
of $B$-decays, both exclusive and inclusive. 
Abundant data in various exclusive channels 
are expected to arrive within the next few years 
from the dedicated $B$-factories BaBar and Belle;  
 their potential impact on our understanding 
of CP-violation at the electroweak scale will crucially depend on our 
possibility to control the effects of strong interactions. 
For exclusive decays 
with only one hadron in the final state, the task is to calculate various 
transition form factors; it has already attracted significant attention
in the literature.     

In this paper we present the first complete results for the  
exclusive semileptonic and rare radiative $B$-decays 
to light vector-mesons in the light-cone sum-rule approach. Exclusive decays, 
which are the principal concern of this work, can be grouped as
semileptonic decays:
\begin{itemize}
 \item $ B_{u,d}\to \rho e \nu $,
 \item $ B_{s}\to K^* e \nu $,
\end{itemize}
rare decays corresponding to $b\to s$ transitions, which we term CKM-allowed: 
\begin{itemize}
 \item $ B_{u,d}\to K^*+\gamma $, ~~~ $ B_{u,d}\to K^*  +l^+l^- $, 
 \item $ B_{s}\to \phi +\gamma $, ~~~ $ B_{s}  \to \phi +l^+l^- $, 
\end{itemize}
and $b\to d$ transitions, which we call CKM-suppressed:
\begin{itemize}
\item 
   $ B_{d}\to (\rho,\omega)+ \gamma $, ~~~ $ B_{d}\to (\rho,\omega) +l^+l^- $, 
\item 
   $ B_{u}\to  \rho + \gamma $, ~~~ $ B_{u}\to \rho + l^+l^- $, 
\item 
$ B_{s}\to K^* +\gamma $, ~~~ $ B_{s}  \to K^* + l^+l^- $. 
\end{itemize}

Let $V$ be
a vector-meson, i.e.\ $\rho$, $\omega$, $K^*$ or $\phi$, and let $p_\mu$,
$\epsilon^*_\mu$ and $m_V$ be its momentum, polarization vector and mass, 
respectively. Let $p_B$ ($m_B$) be the momentum (mass) of the $B$-meson. 
We define
{\em semileptonic} form factors by ($q=p_B-p$)
\begin{eqnarray}
\langle V(p) | (V-A)_\mu | B(p_B)\rangle & = & -i \epsilon^*_\mu (m_B+m_V)
A_1^V(q^2) + i (p_B+p)_\mu (\epsilon^* p_B)\,
\frac{A_2^V(q^2)}{m_B+m_V}\nonumber\\
\lefteqn{+\: i q_\mu (\epsilon^* p_B) \,\frac{2m_V}{q^2}\,
\left(A_3^V(q^2)-A_0^V(q^2)\right) +
\epsilon_{\mu\nu\rho\sigma}\epsilon^{*\nu} p_B^\rho p^\sigma\,
\frac{2V^V(q^2)}{m_B+m_V}\,.}\hspace*{2cm}
\label{eq:SL}
\end{eqnarray}
Note the exact relations
\begin{eqnarray}
 A_3^V(q^2) & = & \frac{m_B+m_V}{2m_V}\, A_1^V(q^2) -
\frac{m_B-m_V}{2m_V}\, A_2^V(q^2),\nonumber\\
A_0^V(0) & = & A_3^V(0), \nonumber\\
\langle V |\partial_\mu A^\mu | B\rangle & = & 2 m_V
(\epsilon^* p_B) A_0^V(q^2).
 \label{eq:A30}
\end{eqnarray}
The second relation in (\ref{eq:A30}) ensures that there is no kinematical
singularity in the matrix element at $q^2=0$.
 
Rare decays are described by the above semileptonic form
factors and the following {\em penguin} form factors:
\begin{eqnarray}
\langle V | \bar \psi \sigma_{\mu\nu} q^\nu (1+\gamma_5) b |
B(p_B)\rangle & = & i\epsilon_{\mu\nu\rho\sigma} \epsilon^{*\nu}
p_B^\rho p^\sigma \, 2 T_1(q^2)\nonumber\\
& & {} + T_2(q^2) \left\{ \epsilon^*_\mu
  (m_B^2-m_{V}^2) - (\epsilon^* p_B) \,(p_B+p)_\mu \right\}\nonumber\\
& & {} + T_3(q^2)
(\epsilon^* p_B) \left\{ q_\mu - \frac{q^2}{m_B^2-m_{V}^2}\, (p_B+p)_\mu
\right\}\label{eq:T}
\end{eqnarray}
with 
\begin{equation}
 T_1(0)  =  T_2(0). \label{eq:T1T2}
\end{equation}
Here $\psi=s,d$. 
All signs are defined in such a way as to render the form factors positive.

The physical range of $q^2$ extends from $q^2_{\rm min} = 0$ to
$q^2_{\rm max} = (m_B-m_V)^2$ for three-body decays and $q^2\equiv 0$
for two-body decays.

The method of light-cone sum-rules was first suggested for the study of 
weak baryon-decays in \cite{BBK} and later extended to heavy-meson decays 
in \cite{chernB}. It is a non-perturbative approach, which combines ideas
of QCD sum-rules \cite{SVZ} 
with the twist-expansion characteristic of hard exclusive 
processes in QCD \cite{exclusive} and makes explicit use of the large 
energy of the final-state vector-meson at small values of the 
momentum-transfer to leptons, $q^2$.  In this respect, the light-cone 
sum-rule approach is complementary to lattice calculations 
\cite{flynn}, which are mainly restricted to form factors at small recoil
(large values of $q^2$). Of course, the light-cone sum-rules lack the 
rigour of the lattice approach. Nevertheless, they prove to provide 
 a powerful non-perturbative model, which is explicitly consistent with 
perturbative QCD and the heavy-quark limit. 

Early studies of exclusive $B$-decays in the light-cone sum-rule
approach were restricted to contributions of leading-twist and 
did not take radiative corrections into account, see 
Refs.~\cite{VMBreview,KRreview} for a review and references 
to original publications. Very recently, these corrections have been 
calculated for the semileptonic $B\to\pi,K e\nu$ decays \cite{BPi}.
In this work we calculate radiative and higher-twist corrections to 
all form factors involving vector-mesons (see above) making use of new 
results on distribution amplitudes of vector-mesons, reported in 
\cite{BBrho,BBKT,BBS}. We find that the corrections in question are fairly
small in all cases.
      
The presentation is organized as follows: in Sec.~2 we recall basic ideas
of the light-cone sum-rule approach  and derive radiative and higher-twist
corrections to the form factors in question in a compact form.
Section~3 presents our main results 
and includes a discussion of input-parameters as well as  error estimates. 
In Sec.~4 we discuss relations between semileptonic and penguin form
factors in the heavy-quark limit. 
Section~5 is
reserved to a summary and conclusions. The paper has two appendices:
in App.~A we collect the relevant loop integrals for the
calculation of radiative corrections. Appendix~B contains a summary 
of the results of \cite{BBrho,BBKT,BBS} on vector-meson 
distribution amplitudes.

\section{Method and Calculation}
\setcounter{equation}{0}

\subsection{General Framework}

Consider semileptonic $B_d\to\rho  e\nu$ and rare   
$B_d\to K^* \ell^+ \ell^-$ decays as representative 
examples.
We choose a $B$-meson ``interpolating current'' $j_B = \bar d i\gamma_5
b$, so that
\begin{equation}\label{eq:deffB}
 \langle 0 | j_B |B(p_B)\rangle = \frac{f_B m_B^2}{m_b}\,,
\end{equation}
where $f_B$ is the usual $B$-decay constant and $m_b$ the $b$-quark mass.
In order to obtain information on the form factors, we study
the  set of suitable correlation functions:\footnote{
In this work we  define invariant functions with respect to the 
Lorentz-structure $\frac{\epsilon^* q}{pq}$ instead of $\epsilon^* q$
\cite{rhoFFs}  
in order to remove a kinematical singularity  for $p\to 0$.} 
\begin{eqnarray}
\lefteqn{i\int d^4y e^{-ip_By} \langle\rho(p)|T(V-A)_\mu(0)
j_B^\dagger(y)|0\rangle\ =\ -i \Gamma^0(p_B^2,q^2) \epsilon^*_\mu}
\hspace*{1.2cm}\nonumber\\
&  &{}  + i \Gamma^+(p_B^2,q^2)\,
\frac{\epsilon^* q}{pq}\, (q+2p)_\mu +
i \Gamma^-(p_B^2,q^2) \,\frac{\epsilon^* q}{pq}\, q_\mu + \Gamma^V(p_B^2,q^2)
\epsilon_\mu^{\phantom{\mu}\alpha\beta\gamma} \epsilon^*_\alpha q_\beta
p_\gamma,\label{eq:CF1}\\
\lefteqn{i\int d^4y e^{-ip_By} \langle K^*(p)|T [\bar s\sigma_{\mu\nu}
\gamma_5 b](0) j_B^\dagger(y)|0\rangle\ =\ {\cal A}(p_B^2,q^2) 
\{\epsilon^*_\mu (2p+q)_\nu - \epsilon^*_\nu
(2p+q)_\mu\}}\hspace*{1.2cm}
\nonumber\\
& & {} - {\cal B}(p_B^2,q^2)\{\epsilon^*_\mu q_\nu - \epsilon^*_\nu
q_\mu)\} - 2
{\cal C}(p_B^2,q^2) \,\frac{\epsilon^* q}{pq}\, \{p_\mu q_\nu - q_\mu p_\nu\}.
\label{eq:CF2}
\end{eqnarray}
The Lorentz-invariant functions $\Gamma^{0,\pm,V},{\cal A},{\cal B},{\cal
C}$
can be calculated in QCD for large Euclidian $p_B^2$. More precisely,
if $m_b^2-p_B^2\ll 0$, then the correlation functions (\ref{eq:CF1}),
(\ref{eq:CF2}) are dominated by the region of small $y^2$ and can be    
systematically expanded in powers of the deviation from the light-cone 
$y^2=0$. The light-cone expansion presents a modification of the usual
Wilson operator product expansion, such that relevant operators are 
non-local and are classified in terms of twist rather than dimension.
Matrix elements of non-local light-cone operators between the vacuum and 
the vector-meson state define meson {\em distribution amplitudes} 
\cite{exclusive}, which describe the partition of the meson-momentum between 
the constituents in the infinite momentum frame. In particular, 
there exist two leading-twist distribution amplitudes for 
vector-mesons, see App.~B, corresponding to longitudinal and transverse
polarizations,
respectively:
\begin{eqnarray}
\langle \rho | \bar u(0) \gamma_\mu d(z) | 0\rangle & = &
f_\rho m_\rho p_\mu \,\frac{\epsilon^* z}{p z}\,
\int_0^1 \!\! du \,e^{i\bar u pz} \,\phi_\parallel(u,\mu),\label{eq:xx}\\
\langle \rho | \bar u(0) \sigma_{\mu\nu} d(z) | 0\rangle & = &
-i f_\rho^T(\mu) (\epsilon^*_\mu p_\nu - p_\mu \epsilon^*_\nu)
\int_0^1 \!\! du \,e^{i\bar u pz} \,\phi_\perp(u,\mu),\label{eq:11}
\end{eqnarray}
and similarly for $K^*$ and $\phi$. 
Here $z$ is an auxiliary light-like vector, 
$u$ is the momentum fraction carried by the 
valence quark, and the decay constants $f_\rho$, $f_\rho^T$ are defined 
in App.~B, while $\mu$ specifies the scale:
extraction of the leading asymptotic behaviour in field theories 
invariably produces singularities, which reflect themselves in the 
scale dependence of distribution amplitudes. As always, this
scale dependence
cancels in physical quantities by a corresponding dependence of coefficient
functions.
 
The invariant amplitudes in (\ref{eq:CF1}), (\ref{eq:CF2}) can be
calculated
in terms of meson distribution amplitudes, in complete analogy with the   
calculation of structure functions in deep inelastic lepton-nucleon
scattering in terms of 
nucleon parton-distributions: the off-shellness $m_b^2-p_B^2$ plays the 
role of photon virtuality $Q^2$.
As an illustration, consider the tree-level  leading-twist result for 
$\Gamma^0$, adapted from Ref.~\cite{rhoFFs}:
\begin{equation}
  \Gamma^0(p_B^2,q^2) = \int_0^1\!\! du\,\frac{1}{m_b^2-up_B^2-\bar u
  q^2}\, f^T_V \phi_\perp(u) \,\frac{m_b^2-q^2}{2u}.
\end{equation} 
We want to emphasize that the procedure is rigorous at this point: 
all corrections can (in principle) be included in a systematic way, 
and their evaluation is precisely the subject of this work.
 
The subtle part concerns the extraction of the $B$-meson contribution
to the invariant amplitudes. The exact amplitude $\Gamma^0$ 
(in nature) has a pole at $p_B^2=m_B^2$ corresponding to the 
intermediate $B$-meson state, and this 
contribution can be written in terms of the form factor $A_1^{B\to\rho}$
defined in (\ref{eq:SL}):
\begin{equation}
   \Gamma^0_{B\mbox{-meson}} =  (m_B+m_\rho) A_1^{B\to\rho}(q^2)\cdot
\frac{1}{m^2_B-p_B^2} \cdot 
        \frac{m_B^2 f_B}{m_b}\,.
\end{equation}
On the other hand, the QCD calculation at $p_B^2 \ll m_b^2 $ is only 
approximate and, continued analytically to ``Minkowskian'' $p_B^2 > m_b^2$,
produces a smooth imaginary part with no sign of a pole behaviour.
To proceed,  we invoke the concept of {\em duality}, assuming that 
the exact spectral density and the one calculated in QCD coincide 
{\em on the average}, that is integrated over a sufficient region of
energy. 
In particular, we assume that the $B$-meson contribution is obtained 
by the integral of the QCD spectral density over the {\em duality region}:
\begin{equation}
   \Gamma^0_{B\mbox{-meson}} = \frac{1}{2\pi
i}\int\limits_{m_b^2}^{s_0}\frac{ds}{s-p_B^2}
   \,{\rm Disc\,}_{p_B^2}\Gamma^0_{\rm QCD}(s,q^2).
\label{eq:disprel}
\end{equation}
The parameter $s_0\approx(34$--$35)$~GeV$^2$ is called ``continuum-threshold'' 
and is fixed from QCD sum-rules for $f_B$, see e.g.\ \cite{bestseller}.
Equating the above  two representations, one obtains a {\em light-cone
sum-rule} for the form factor $A_1$. Sum-rules for the other form factors
are constructed in precisely the same manner.

While the accuracy of the QCD calculation can be controlled (and improved), 
the duality approximation introduces an irreducible uncertainty in 
predictions for the form factors, which is usually believed to be of 
order (10--15)\%. 
Practical calculations in the sum-rule framework involve some technical
tricks to reduce this uncertainty, e.g.\ Borel transformation, which we 
will not discuss here. These techniques are well established and 
their detailed description in the particular context of light-cone
sum-rules can be found, for instance, in Refs.~\cite{KRreview,rhoFFs}. 
The work \cite{rhoFFs} 
also contains a detailed comparison of the light-cone sum-rule approach to
traditional QCD sum-rules and can serve as introduction for  
the  more theoretically-minded reader.


\subsection{Radiative Corrections}

Radiative corrections to the sum-rules correspond to one-loop 
corrections to the coefficient functions in front of leading-twist
distribution amplitudes and are given by the diagrams shown in Fig.~1.
%
\begin{figure}
\vspace*{-3.5cm}
\centerline{\epsfysize=0.98\textheight\epsfbox{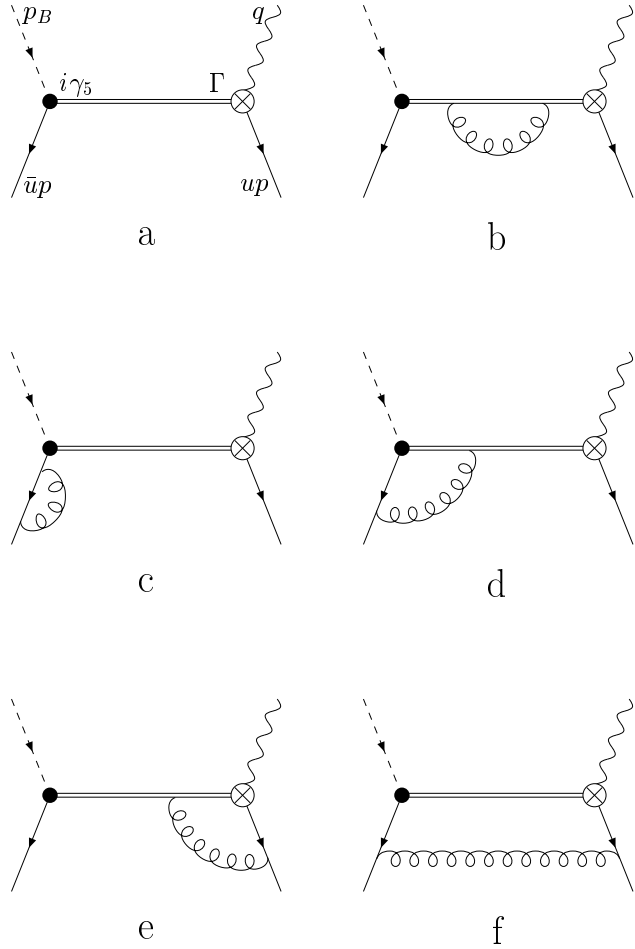}\vspace*{-5cm}}
\caption[]{The leading-order diagram (a) and one-loop radiative corrections
  (b--f).}\label{fig:diagrams}
\end{figure}
%
The calculation is done in dimensional regularization, and it is 
sufficient to consider matrix elements over on-shell massless 
quark and antiquark carrying momentum fraction $up$ and $\bar u p$,
respectively. The transversely polarized and longitudinally 
polarized meson states are projected on by 
\begin{eqnarray}
\langle V_\perp(p)|\bar u_a(0) d_b(x) |0\rangle 
& = & -\frac{i}{4}\, f_V^T [\sigma_{\mu\nu}]_{ba}
\epsilon^{*\mu} p^\nu \int du\, e^{i\bar u px} \phi_\perp(u)
\label{eq:proj1}\\
& \equiv & -\frac{1}{8}\, f_V^T [\sigma^{\mu\nu}\gamma_5]_{ba}
\epsilon_{\mu\nu\rho\sigma} \epsilon^{*\rho} p^\sigma
 \int du\, e^{i\bar u px} \phi_\perp(u),
\label{eq:proj2}\\
\langle V_\parallel(p)|\bar u_a(0) d_b(x) |0\rangle 
& \equiv & \frac{1}{4}\,f_V m_V [\slash{p}]_{ba}
\,\frac{\epsilon^* x}{px}\, \int du\, e^{i\bar u px} \phi_\parallel(u)
\nonumber\\&&
{}\stackrel{m_V^2\to 0}{\longrightarrow} \frac{1}{4}\,f_V 
[\slash{p}]_{ba} \int du\, e^{i\bar u px} \phi_\parallel(u),
\label{eq:rhoparallel}
\end{eqnarray}
where $a,b$ are spinor indices.
In the last line in (\ref{eq:rhoparallel}) we made use of the fact that for
ultrarelativistic longitudinal vector-mesons  $\epsilon_\mu
\to p_\mu/m_V$ up to $O(m_V^2/|\vec{p}|^2)$ corrections. 
This is a justified approximation for the calculation of 
radiative corrections to leading-twist accuracy, to which end 
the meson-mass can be neglected throughout. 
For further use we introduce the notation for the projection operators:
\begin{eqnarray}
{\cal P}_\parallel &=& \frac{1}{4} f_V \slash{p},
\nonumber\\
{\cal P}_{\perp} &=& -\frac{i}{4}f_V^T
               \sigma_{\alpha\beta}\epsilon^{*\alpha}p^\beta
\quad {\rm \ or\ } \quad {\cal P}_{\perp}^{(5)} = -\frac{1}{8}f_V^T
\sigma_{\alpha\beta}\gamma_5\epsilon^{\alpha\beta\rho\sigma}\epsilon^*_{\rho}
p_\sigma.
\label{eq:projectors}
\end{eqnarray}
In what follows, they will be treated as $D$-dimensional objects . 

The calculation in question is in principle straightforward and similar 
to the existing calculations of NLO corrections to hard exclusive
processes \cite{piEM}. 
One has to consider one-loop diagrams 
with a heavy-quark and two different kinematic invariants $q^2$ and $p_B^2$, 
which makes formulas rather cumbersome.
The specific requirement is to organize the expressions 
in a form suitable for a dispersion representation in $p_B^2$, cf.\ 
Eq.~(\ref{eq:disprel}), so that continuum subtraction can be made. 

Analytic expressions for $B$-decays to 
light pseudoscalar-mesons $\pi,K$  have been made available recently 
\cite{BPi}. For vector-mesons the 
number of form factors is so large that working out (relatively) 
compact analytic expressions is not worth the effort. In this work we 
prefer to give the formulae in terms of traces and general momentum
integrals (see below and App.~A), which can be compiled and evaluated 
numerically using the {\sc mathematica} programming language\footnote{
The computer code is available from P.B. upon request.}.

A usual subtlety concerns the treatment of $\gamma_5$. The results for the
form factors given below are obtained using  ``naive dimensional 
regularization'' (NDR), and the same scheme has to be applied to the 
calculation of Wilson coefficients for penguin operators.  

There are two form factors in whose calculation one
encounters an odd number of $\gamma_5$ in traces, which could cause
ambiguities: $V$ and $T_1$. Only transverse mesons contribute to these
form factors. In both cases, a possible ambiguity comes solely from 
the $B$-vertex correction in Fig.~1d, whereas in all other diagrams 
contraction of $\gamma$ matrices over $\gamma_5$ can be avoided. 
There are several ways out:
(a) use a 't Hooft-Veltman prescription for $\gamma_5$ and apply a 
   finite renormalization to restore the Ward identities, as in \cite{larin};
(b) instead of the ``natural'' projection (\ref{eq:proj1}),
   use (\ref{eq:proj2}), which introduces a second $\gamma_5$ and
thus eliminates the problem;
(c) modify the definition of the form factors (\ref{eq:T}) to
\begin{eqnarray}
\langle V|\bar s \sigma_{\mu\nu}\gamma_5 b | B\rangle & = &
A(q^2) \left\{\epsilon^*_\mu (p_B+p)_\nu - (p_B+p)_\mu
  \epsilon^*_\nu\right\} - B(q^2)
\left\{\epsilon^*_\mu q_\nu - q_\mu \epsilon^*_\nu\right\}\nonumber\\
& & {} -  C(q^2) \,\frac{\epsilon^*
  p_B}{m_B^2-m_{V}^2} \,\left\{ (p_B+p)_\mu q_\nu - q_\mu
  (p_B+p)_\nu \right\}.
\label{eq:ABC}
\end{eqnarray}
Using 
$$
\sigma_{\mu\nu}\gamma_5 = -\frac{i}{2}\,\epsilon_{\mu\nu\rho\sigma}
\,\sigma^{\rho\sigma}\label{eq:sigma}
$$
and contracting with $q^\nu$, one finds
\begin{eqnarray}
A(q^2) &=& T_1(q^2),
\nonumber\\
B(q^2) &=& \frac{m_B^2-m_V^2}{q^2}\left[T_1(q^2)-T_2(q^2)\right],
\nonumber\\
C(q^2) &=& T_3(q^2) - \frac{m_B^2-m_V^2}{q^2}\left[T_1(q^2)-T_2(q^2)\right],
\label{eq:eq}
\end{eqnarray}
from which the relation (\ref{eq:T1T2}) follows. It is thus sufficient to
calculate $A$, $B$ and $C$ instead of $T_i$ with the premium to avoid
any  $\gamma_5$ problem. We have checked that all of the above 
prescriptions yield identical results.

After these preliminary remarks, we are now in a position to calculate 
the diagrams in Fig.~1. The tree-level contribution of Fig.~1a equals
\begin{equation}
T^{(0)} = \frac{i}{s}\, {\rm Tr}(\Gamma (\slash{p}_B-\bar u \slash{p} +
m_b)\gamma_5 {\cal P}),
\label{eq:tree}
\end{equation}
where $\Gamma$ is the Dirac structure of the weak vertex, ${\cal P}$ is 
one of the projection operators defined in Eqs.~(\ref{eq:projectors}),
and 
$$s=m_b^2-u p_B^2-\bar u q^2.$$
It proves convenient to replace in Eq.~(\ref{eq:tree})
the running $\overline{\rm MS}$ $b$-quark
mass by the one-loop pole mass, which is given by
\begin{equation}
m_{\rm pole} = m_{\overline{\rm MS}} \left\{ 1 + C_F\,\frac{g^2}{4\pi^2}\left(
    1-\frac{3}{4}\, \ln\,\frac{m^2}{\mu^2}\right)\right\}.
\end{equation}
This replacement induces the radiative correction
\begin{eqnarray}
T^{\rm pole} & = & 2i\,\frac{m_b^2}{s^2} \,C_F\,\frac{g^2}{4\pi^2}
\,\left(1-\frac{3}{4}\, \ln\,\frac{m_b^2}{\mu^2}\right) {\rm Tr}({\cal P}
\Gamma(\slash{p}_B - \bar u \slash{p} + m_b)\gamma_5)\nonumber\\
& &{} - \frac{i}{s}\, m_b C_F\,
\frac{g^2}{4\pi^2} \left(1-\frac{3}{4}\,
  \ln\,\frac{m_b^2}{\mu^2}\right) {\rm Tr}({\cal P}\Gamma (\slash{p}_B-\bar 
u\slash{p})\gamma_5).
\end{eqnarray}

The general strategy is to simplify the traces as much as
possible, but to keep ${\cal P}$ and $\Gamma$ arbitrary. Also contraction
of $\gamma$ matrices over $\gamma_5$ is only  allowed in
the $B$-vertex correction.

It turns out that all one-loop diagrams  can be expressed in
terms of the following traces:
\begin{eqnarray}
\,{\rm Tr}_1 & = & {\rm Tr}\left({\cal P}\Gamma\slash{q}\gamma_5\right) \equiv
{\rm Tr}\left({\cal P}\Gamma\slash{p}_B\gamma_5\right),\nonumber\\
\,{\rm Tr}_2 & = & {\rm Tr}\left({\cal P}\Gamma\gamma_5\right),\nonumber\\
\,{\rm Tr}_3 & = & {\rm Tr}\left({\cal P}\slash{q}\Gamma\gamma_5\right) \equiv
{\rm Tr}\left({\cal P}\slash{p}_B\Gamma\gamma_5\right),\nonumber\\
\,{\rm Tr}_4 & = & {\rm Tr}({\cal P}\slash{q}\Gamma\slash{q}\gamma_5).
\end{eqnarray}
Let us also introduce
\begin{equation}
a_{\cal P} {\cal P} := \gamma_\alpha {\cal P} \gamma^\alpha,\qquad 
a_\Gamma \Gamma :=  \gamma_\alpha \Gamma \gamma^\alpha.
\end{equation}
The $b$-quark self-energy diagram in Fig.~1b is:
\begin{eqnarray}
T^{\rm SE}_b & = & -\frac{g^2C_F}{s^2}\,\left\{ \left[ 4 m_b^2
\left(Y+(1-\epsilon) Z\right) + 2 s (1-\epsilon)(Y-Z)
\right] (\,{\rm Tr}_1+m_b \,{\rm Tr}_2)\right.\nonumber\\
& & \left. -2m_b s \left(Y+(1-\epsilon)Z\right)\,{\rm Tr}_2\right\},
\end{eqnarray}
where $D=4-2\epsilon$ and the expressions for momentum integrals $Y,Z$
are given in App.~A. The self-energy insertions in external light-quark
legs in Fig.~1c only contribute logarithmic terms in dimensional
regularization:
\begin{equation}
T^{\rm SE}_l = \frac{g^2 C_F}{s} \left({\rm Tr}_1+m_b {\rm
    Tr}_2\right)  \left(\ln \frac{m_b^2}{\mu_{UV}^2}-
\ln \frac{m_b^2}{\mu_{IR}^2}\right),
\end{equation}
where we distinguish between the ultra-violet scale $\mu_{UV}$,
which is to be identified with the renormalization scale of the curent
$j_B$ and the penguin operators, and the 
infra-red renormalization-scale $\mu_{IR}$ corresponding to the
factorization scale in meson distribution amplitudes. 

For the $B$-vertex correction in Fig.~1d, one obtains:
\begin{equation}
T^B =  2\,\frac{g^2C_F}{s}\left\{ \left(-8\bar
C(1-\epsilon)-1-m_b^2\bar B + \bar u (p_B^2-q^2) \bar A\right)
(\,{\rm Tr}_1+m_b \,{\rm Tr}_2) -m_b s\bar B \,{\rm Tr}_2\right\}.
\end{equation}
For the weak vertex in Fig.~1e we find:
\begin{eqnarray}
T^W & = & {}-\frac{g^2C_F}{s}\left\{ a_\Gamma^2 \, C \,(\,{\rm Tr}_1 + m_b
\,{\rm Tr}_2) + a_\Gamma\, D \,(q^2 \,{\rm Tr}_3 + 
m_b \,{\rm Tr}_4) + (p_B^2 - q^2) u
a_\Gamma\, E \,{\rm Tr}_3\right.\nonumber\\
\lefteqn{\left. + 2 (p_B^2-q^2) u A (\,{\rm Tr}_1 + m_b \,{\rm Tr}_2)  + m_b B
\left[ 2 (m_b \,{\rm Tr}_1 + q^2 \,{\rm Tr}_2) - a_\Gamma (m_b \,{\rm Tr}_3 +
\,{\rm Tr}_4)\right] \right\}.}\hspace*{0.4cm}
\end{eqnarray}
Finally, the box-diagram in Fig.~1f can be written as
\begin{equation}
T^{\rm box} = -g^2 C_F a_{\cal P} \left\{ a_{\cal P} H 
(\,{\rm Tr}_1 + m_b \,{\rm Tr}_2) +
I (-m_b \,{\rm Tr}_4 + (s-m_b^2) \,{\rm Tr}_3) + 
 {B}_{u=1} \,{\rm Tr}_3\right\},
\end{equation}
where ${B}_{u=1}$ is the limiting value of $B$ for $u \to 1$.

Definitions and explicit expressions for the  one-loop integrals 
$A$, $B$, $C$, etc.,  are given in App.~A.


\subsection{Higher-Twist Contributions}

Higher-twist terms generically refer to contributions 
to the light-cone expansion of the
correlation functions (\ref{eq:CF1}) and (\ref{eq:CF2}), which are
suppressed by powers of $1/(m_b^2-p_B^2)$. In the sum-rules, such 
corrections are suppressed by powers of the Borel parameter. 
Higher-twist corrections are usually divided into ``kinematical'',
originating from the non-zero mass of the vector-meson, and ``dynamical'',
related to contributions of higher Fock-states and transverse
quark-motion. In this paper we take into account both effects to twist-4
accuracy, making use of the new results on distribution amplitudes 
of vector-mesons reported in \cite{BBKT,BBS} and summarized in 
App.~B.

%
\begin{figure}[t]
\vspace*{-3.5cm}
\centerline{\epsfysize=\textheight\epsfbox{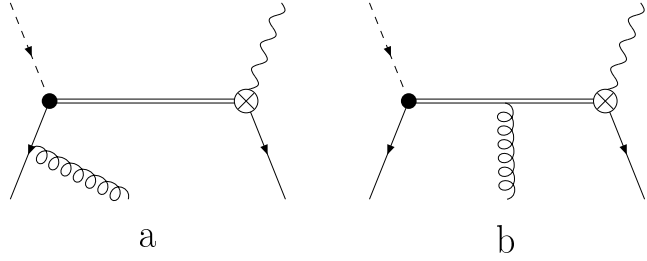}\vspace*{-14cm}}
\caption[]{The higher-twist contributions.}\label{fig:twist}
\end{figure}
%

The calculation is most conveniently done using the background-field
approach of \cite{BB89}. The diagrams of the type shown in Fig.~2a
are taken into account within this method by the expansion
of the non-local quark-antiquark operator in powers of the deviation
from the light-cone; they give rise to contributions of two-particle
distribution amplitudes of higher-twist, see Eqs.~(\ref{eq:OPE1}) and 
(\ref{eq:OPE2}). The contribution of the gluon-emission from the heavy-quark
is calculated using the light-cone expansion of the quark-propagator
\cite{BB89,Bel95}:
\begin{eqnarray}
\lefteqn{\langle 0 | T \{b(x) \bar b(0)\} | 0 \rangle\ =}
\nonumber\\
& = & {} i S_b(x) - ig
\int\!\! \frac{d^4k}{(2\pi)^4} \,e^{-ikx}\!\! \int_0^1 \!\! dv\left[
\frac{1}{2} \,\frac{\slash{k}+m_b}{(m_b^2-k^2)^2}
\,G^{\mu\nu}(vx)\sigma_{\mu\nu} + \frac{v}{m_b^2-k^2} \,x_\mu
G^{\mu\nu}(vx) \gamma_\nu\right]\!,\makebox[8mm]{\ }
\end{eqnarray}
where $S_b(x)$ is the free quark-propagator. 
As in the case of radiative corrections, our strategy in this work
is to derive the most general expression for all form factors in question,
suitable for implementation in  analytic/numerical calculations using
{\sc mathematica}. We obtain
\begin{eqnarray}
{\rm CF} & = & \frac{1}{4}\int_0^1\!\! du \left\{ i f_Vm_V \left[
\left( \Phi_\parallel^{(i)}(u)\epsilon^*_\alpha \,\frac{\partial}{\partial
Q_\alpha} + \frac{\epsilon^* q}{pq} \,\frac{1}{16}\, m_V^2
{\Bbb A}(u)\,\frac{\partial^2}{\partial Q_\rho\partial Q^\rho} \right)
Tr(\Gamma S_b(Q) \gamma_5 \slash{p})\right.\right.
\nonumber\\
& & {} -g_\perp^{(v)}(u) Tr(\Gamma S_b(Q)\gamma_5 \slash{\epsilon}^*) -
\frac{\epsilon^* q}{pq} \, \frac{1}{2}\, m_V^2 {\Bbb C}^{(i)}(u)
\,\frac{\partial}{\partial Q_\alpha}\, Tr(\Gamma
S_b(Q)\gamma_5\gamma_\alpha)
\nonumber\\
& & \left. - \frac{i}{4}\,
\epsilon_{\alpha\beta\gamma\delta} \epsilon^{*\beta}p^\gamma
g_\perp^{(a)}(u) \,\frac{\partial}{\partial Q_\delta} Tr(\Gamma S_b(Q)
\gamma_\alpha)\right] - f_V^T \left[ \left( \phi_\perp(u) -
\frac{1}{16}\, m_V^2 {\Bbb A}_T(u) \,\frac{\partial^2}{\partial Q_\rho
\partial Q^\rho}\right)\right.
\nonumber\\
& & {}\times Tr(\Gamma S_b(Q) \gamma_5
\sigma_{\alpha\beta}) \epsilon^{*\alpha} p^\beta - \frac{\epsilon^* q}{pq}
\, m_V^2 {\Bbb B}_T^{(i)}(u)  
p^\alpha \,\frac{\partial}{\partial Q_\beta} Tr(\Gamma S_b(Q)
\gamma_5 \sigma_{\alpha\beta}) 
\nonumber\\
& & {}-\frac{i}{2}\,\left( 1 - \frac{m_q+m_{\bar q}}{m_V}\,
\frac{f_V}{f_V^T} \right) m_V^2 h_\parallel^{(s)}(u) \epsilon^{*\alpha}
\,\frac{\partial}{\partial Q_\alpha} \,Tr(\Gamma S_b(Q)\gamma_5)
\nonumber\\
& & \left.\left. {} - \frac{1}{2}\, m_V^2 {\Bbb C}_T^{(i)}(u)
\epsilon^{*\alpha}\,\frac{\partial}{\partial Q_\beta}\, Tr(\Gamma S_b(Q)
\gamma_5\sigma_{\alpha\beta} ) \right]\right\}
\nonumber\\
& & + \frac{i}{4}f_V m_V \int_0^1\!\! dv \!\int\!\! {\cal D}\alpha
\,\Big[m_b^2 - ( q + (\alpha_1 + v \alpha_3) p)^2\Big]^{-2}
\Bigg[ 2 v (pq) \left( {\cal A}(\underline{\alpha}) + 
{\cal V}(\underline{\alpha}) \right)
Tr(\Gamma\slash{\epsilon}^*\slash{p}\gamma_5)\,
\nonumber\\
& & {}+ m_V^2\, \frac{\epsilon^* q}{pq}\, \left( 2
  \Phi(\underline{\alpha}) + \Psi(\underline{\alpha}) - 2
\widetilde{\Phi}(\underline{\alpha}) -
\widetilde{\Psi}(\underline{\alpha}) 
\right)
Tr(\Gamma(\slash{q}+m_b) \slash{p}\gamma_5)
\nonumber\\
& &  + 4 m_V^2 v (\epsilon^* q) \left(
\widetilde{\Phi}(\underline{\alpha}) - \Phi(\underline{\alpha})\right) 
Tr(\Gamma\gamma_5) -
m_V^2 v\, \frac{\epsilon^* q}{pq} \, \Psi(\underline{\alpha})
Tr(\Gamma\slash{q}\slash{p}\gamma_5) \Bigg]
\nonumber\\
& & + \frac{i}{4}f_V^T m_V^2 \int_0^1\!\! dv \!\int\!\! {\cal D}\alpha
\,\Big[m_b^2 - ( q + (\alpha_1 + v \alpha_3) p)^2\Big]^{-2}
\Bigg[ -2v(\epsilon^* q) {\cal T}(\underline{\alpha})
Tr(\Gamma\slash{p}\gamma_5)
\nonumber\\
& & {} + \left( {\cal S}(\underline{\alpha}) - \widetilde{\cal
S}(\underline{\alpha}) +T_1^{(4)}(\underline{\alpha}) - T
_2^{(4)}(\underline{\alpha}) + T_3^{(4)}(\underline{\alpha}) -
T_4^{(4)}(\underline{\alpha}) \right) Tr(\Gamma(\slash{q}+m_b)
\slash{\epsilon}^*
\slash{p} \gamma_5)
\nonumber\\
& & {} + 2 v \left(T_2^{(4)}(\underline{\alpha}) - 
T_4^{(4)}(\underline{\alpha}) -
{\cal S}(\underline{\alpha}) - \widetilde{\cal
  S}(\underline{\alpha})\right) 
\Big[ (\epsilon^* q)
Tr(\Gamma\slash{p}\gamma_5) -(pq) Tr(\Gamma\slash{\epsilon}^*
\gamma_5) \Big]
\nonumber\\
& & {} + 2 v \left( T_3^{(4)}(\underline{\alpha}) - 
T_4^{(4)}(\underline{\alpha}) -
\widetilde{\cal S}(\underline{\alpha})\right)
Tr(\Gamma\slash{q}\slash{p}\slash{\epsilon}^* \gamma_5)\Bigg],
\label{eq:monster}
\end{eqnarray}
where $Q = q + \bar u p$ and 
${\rm CF} \in \{\Gamma^{0,\pm,V},{\cal A},{\cal B},{\cal
C}\}$. Definitions and explicit expressions for the numerous distribution
amplitudes are collected in App.~B \footnote{Despite appearance, the 
number of non-perturbative parameters in the description of higher-twist
distributions is small, since they are related by exact equations 
of motion, see \cite{BBKT,BBS} and App.~B.}. 
In addition, we use the notation
\begin{equation}
\Phi_\parallel^{(i)}(u) = 
\int_0^u\! dv\, \Big(\phi_\parallel(v)-g_\perp^{(v)}(v)\Big).
\end{equation}
To leading-twist
accuracy, our result agrees with the expressions available in the 
literature, see \cite{ABS,rhoFFs,aliev}\footnote{
 The sum-rule for $T_1$ given in 
\cite{ABS,aliev} misses a contribution of $\Phi_\parallel$; this term 
can be formally viewed as part of the kinematic higher-twist correction,
which is included in \cite{ABS,aliev} only partially.}.  

\section{Results}
\setcounter{equation}{0}

In this section we present results of the numerical analysis of the
light-cone sum-rules for the form factors defined in 
(\ref{eq:SL}) and (\ref{eq:T}) for $B$- and $B_s$-decays. The sum-rules
depend on several parameters, those needed to describe the $B$-meson
on the one hand and those describing the vector-meson on the other
hand. The former  are essentially $f_B$ ($f_{B_s}$), the leptonic
decay constant defined in (\ref{eq:deffB}), the $b$-quark mass $m_b$,
the continuum-threshold $s_0$ introduced in (\ref{eq:disprel}), and the
Borel parameter $M^2$ mentioned in Sec.~2.1. Lacking experimental
determination of $f_B$ and $f_{B_s}$, we determine their values from
two-point QCD sum-rules to $O(\alpha_s)$ accuracy (see e.g.\ 
\cite{bestseller}); this fixes $s_0$, which depends on $m_b$, and also the
``window'' in $M^2$ in which the sum-rules are evaluated. We then use
the same values for $m_b$, $s_0$ and $M^2$ in both
the QCD sum-rule for $f_B$ and the light-cone sum-rules for the form
factors,\footnote{To be precise, the expansion-parameter of the
  light-cone correlation function is $uM^2$ rather than $M^2$. Because of
  this, in the light-cone sum-rules we use an ``effective'' Borel
  parameter $M_{\rm eff}^2$ defined by $\langle u\rangle M_{\rm eff}^2
  \equiv M^2_{2pt}$, $M^2_{2pt}$ being the Borel parameter used in the
  QCD sum-rules for $f_B$.} 
which helps to reduce the systematic uncertainty of the
approach. The corresponding parameter-sets and results for the decay
constants are given in Table~\ref{tab:fB}. The question of the value 
of the $b$-quark mass has attracted considerable attention recently; 
following these developments \cite{mb}, we use the value
$m_b=(4.8\pm 0.1)\,$GeV. Our results for $f_B$ agree well with
new lattice determinations \cite{fBlattice}.

\begin{figure}
\centerline{\epsffile{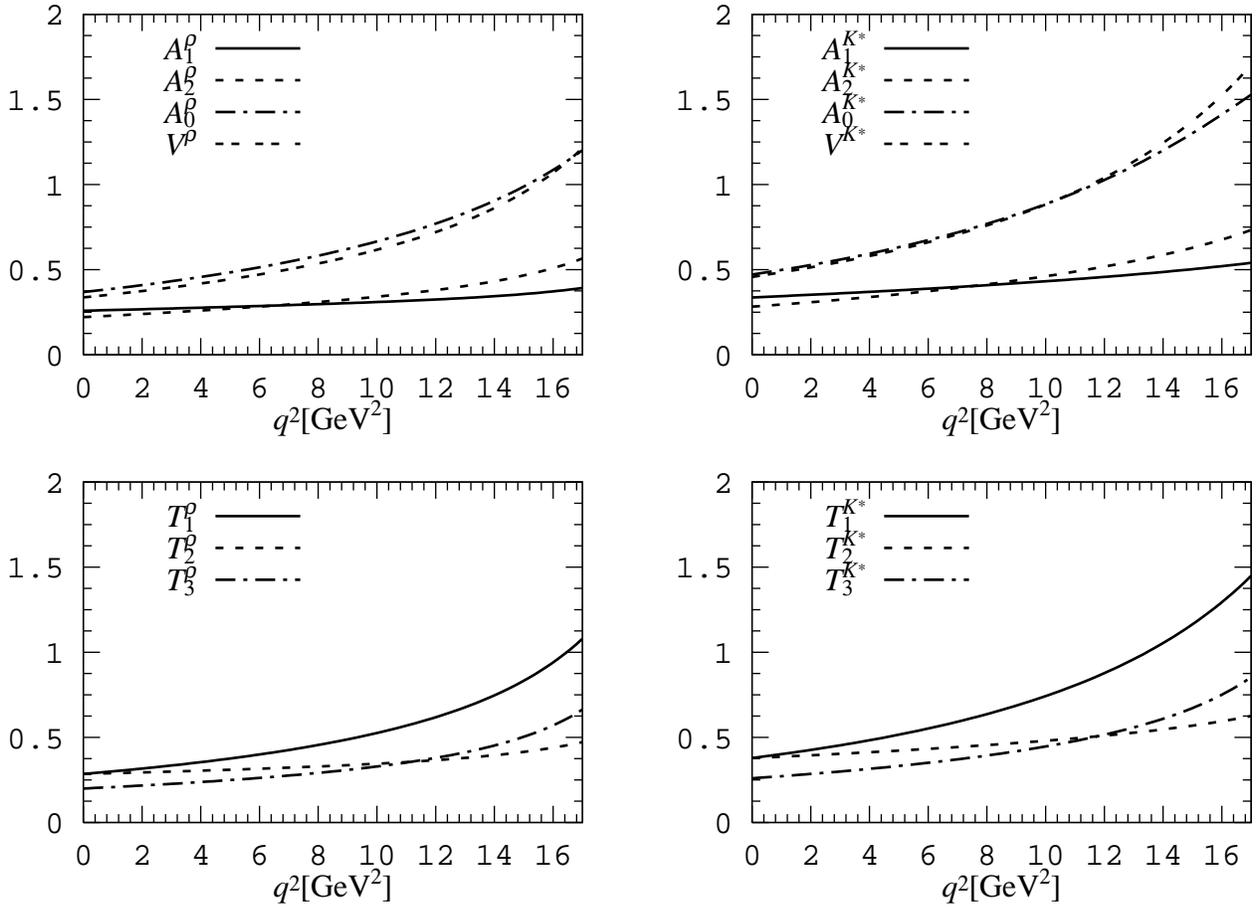}}
\caption[]{Light-cone sum-rule results for $B_{u,d}\to\,$vector-meson form
factors. Renormalization scale for $T_i$ is $\mu = m_b =
4.8\,$GeV. Further parameters: $m_b = 4.8\,$GeV, $s_0 = 33.5\,$GeV$^2$,
$M^2 = 6\,$GeV$^2$.}\label{fig:B}
\end{figure}

\begin{figure}
\centerline{\epsffile{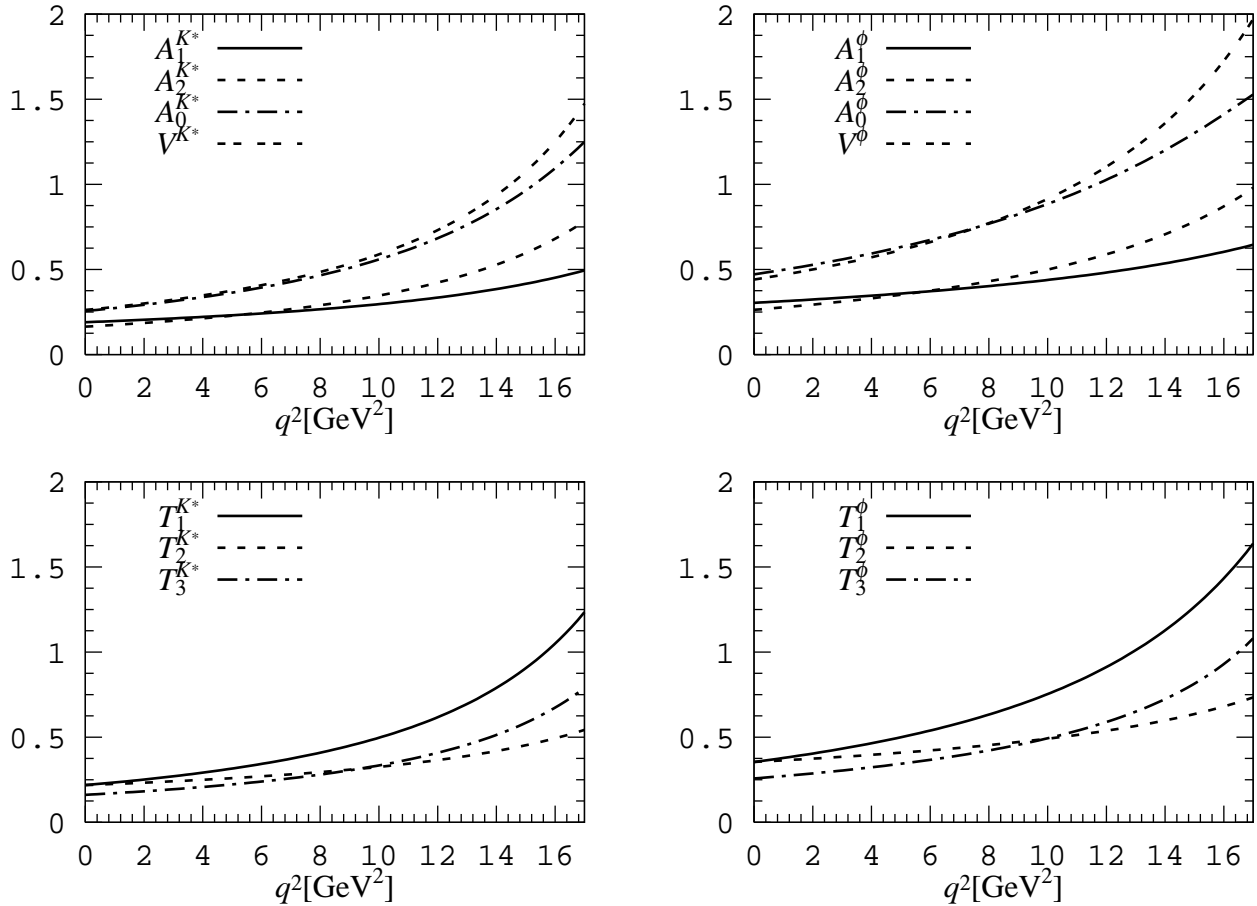}}
\caption[]{Light-cone sum-rule results for $B_s\to\,$vector-meson form
factors. Renormalization scale for $T_i$ is $\mu = m_b =
4.8\,$GeV. Further parameters: $m_b = 4.8\,$GeV, $s_0 = 34.5\,$GeV$^2$,
$M^2 = 6\,$GeV$^2$.}\label{fig:Bs}
\end{figure}

The parameters of light mesons are collected in
App.~B, Tables~\ref{tab:para} and \ref{tab:para2}. These parameters are
evaluated at the factorization scale $\mu_{IR}^2 = m_B^2-m_b^2 =
4.8\,$GeV$^2$, which is the typical virtuality of the virtual 
$b$-quark in the process. The penguin form factors also depend on the
ultra-violet renormalization scale of the effective weak Hamiltonian,
for which we choose $\mu_{UV}=m_b$. Using the central values of all
parameters, we obtain the form factors plotted in Figs.~\ref{fig:B}
and \ref{fig:Bs}. For their representation in algebraic form, a
parametrization in terms of three parameters proves convenient:
\begin{equation}
F(q^2) = \frac{F(0)}{1-a_F\,\frac{q^2}{m_B^2} + b_F \left
    ( \frac{q^2}{m_B^2} \right)^2},
\end{equation}
with the fit parameters $F(0)$, $a_F$ and $b_F$. Here $m_B$ is the mass
of the relevant $B$-meson, i.e.\ $m_{B_{u,d}}$ for $B_{u,d}$-decays
and 
$m_{B_s}$ for
$B_{s}$-decays. This parametrization describes all 28
form factors to an accuracy of 1.8\% or better for $0\leq q^2\leq
17\,$GeV$^2$.

\begin{table}
\renewcommand{\arraystretch}{1.4}
\addtolength{\arraycolsep}{3pt}
$$
\begin{array}{|l|lll|}\hline
m_b\,[{\rm GeV}] & 4.7 & 4.8 & 4.9\\ \hline
s_0\,[{\rm GeV}^2] & 34.5\pm 0.5 & 33.5\pm 0.5 & 32.5\pm 0.5\\
f_B\,[{\rm MeV}] & 177\pm 5 & 150\pm 5 & 123 \pm 5\\ \hline 
s_0\,[{\rm GeV}^2] & 35.5\pm 0.5 & 34.5\pm 0.5 & 33.5\pm 0.5\\
f_{B_s}\,[{\rm MeV}] & 191\pm 5 & 162\pm 5 & 135 \pm 5\\ \hline 
\end{array}
$$
\renewcommand{\arraystretch}{1}
\addtolength{\arraycolsep}{-3pt}
\caption[]{Values for $f_B$ and $f_{B_s}$ from QCD sum-rules in
  dependent on the $b$-quark mass. The Borel parameter window is $M^2
  = (4$--$8)\,$GeV$^2$.}\label{tab:fB}
\end{table}

Let us now discuss the dependence of the results on the
input-parameters and approximations involved. First we note that the net
impact of radiative corrections is very small at small $q^2$ and at
most 5\% at $q^2=0$. Their effect increases, however, at large $q^2$
and leads to a decrease of the form factors $A_2$ and $T_3$ at
$q^2=17\,$GeV$^2$ by 20\%
with respect to their tree-level values; the
impact on the other form factors stays in the 5\% range. 
The small effect of radiative corrections was
anticipated in the tree-level analysis of Ref.~\cite{rhoFFs} and also
observed in the calculation of $O(\alpha_s)$ corrections for
$B\to\,$pseudoscalar decays \cite{BPi}.
It is due to the fact that the biggest contribution to radiative
corrections (in Feynman gauge) comes from the $B$-vertex correction
diagram, which enters both the calculation of $f_B$ and the light-cone
correlation functions and cancels in the ratio that gives the form
factors. Although literally we only calculated radiative corrections
to the leading-twist contribution to the light-cone expansion, it is unlikely
that yet unknown corrections to the higher-twist terms could
change this pattern dramatically. We thus believe that radiative
corrections are under good control. 

\begin{table}
\renewcommand{\arraystretch}{1.5}
\addtolength{\arraycolsep}{3pt}
$$
\begin{array}{|l|ccc|ccc|l|}
\hline
& F(0) & a_F & b_F & F(0) & a_F & b_F & \\ \hline
A_1^{\rho} & 0.261 & 0.29 & -0.415& 0.337& 0.60 & -0.023& A_1^{K^*} \\
A_2^{\rho} & 0.223 & 0.93 & -0.092 & 0.283 & 1.18 & \phantom{-}0.281
& A_2^{K^*}\\
A_0^\rho & 0.372 & 1.40 & \phantom{-}0.437 & 0.470 & 1.55 &
\phantom{-}0.680 & A_0^{K^*} \\
V^\rho & 0.338 & 1.37 & \phantom{-}0.315
 & 0.458 & 1.55 & \phantom{-}0.575 & V^{K^*} \\ \hline
T_1^\rho & 0.285 & 1.41 & \phantom{-}0.361
 & 0.379 & 1.59 & \phantom{-}0.615 & T_1^{K^*}\\
T_2^\rho & 0.285 & 0.28 & -0.500
 & 0.379 & 0.49 & -0.241 & T_2^{K^*}\\
T_3^\rho & 0.202 & 1.06 & -0.076
 & 0.261 & 1.20 & \phantom{-}0.098 & T_3^{K^*}\\
\hline
\end{array}
$$
\caption{$B_{u,d}$-decay form factors in a three-parameter fit. 
Renormalization scale
for $T_i$ is $\mu = m_b = 4.8\,$GeV. The theoretical uncertainty is
estimated as 15\%.}\label{tab:fit}
$$
\begin{array}{|l|ccc|ccc|l|}
\hline
& F(0) & a_F & b_F & F(0) & a_F & b_F & \\ \hline
A_1^{K^*} & 0.190 & 1.02 & -0.037& 0.296 & 0.87 & -0.061 & 
A_1^{\phi} \\
A_2^{K^*} & 0.164 & 1.77 & \phantom{-}0.729 & 0.255 & 1.55 & 
\phantom{-}0.513 & A_2^{\phi}\\
A_0^{K^*} & 0.254 & 1.87 & \phantom{-}0.887 & 0.382 & 1.77 &
\phantom{-}0.856 & A_0^{\phi} \\
V^{K^*} & 0.262 & 1.89 & \phantom{-}0.846
 & 0.433 & 1.75 & \phantom{-}0.736 & V^{\phi} \\ \hline
T_1^{K^*} & 0.219 & 1.93 & \phantom{-}0.904
 & 0.348 & 1.82 & \phantom{-}0.825 & T_1^{\phi}\\
T_2^{K^*} & 0.219 & 0.85 & -0.271
 & 0.348 & 0.70 & -0.315 & T_2^{\phi}\\
T_3^{K^*} & 0.161 & 1.69 & \phantom{-}0.579
 & 0.254 & 1.52 & \phantom{-}0.377 & T_3^{\phi}\\
\hline
\end{array}
$$
\caption{$B_s$-decay form factors in a three-parameter fit. 
Renormalization scale
for $T_i$ is $\mu = m_b = 4.8\,$GeV. The theoretical uncertainty is
estimated as 15\%.}\label{tab:fits}
\renewcommand{\arraystretch}{1}
\addtolength{\arraycolsep}{-3pt}
\end{table}

The next question concerns the convergence of the light-cone
expansion. The higher-twist terms have several sources: some depend
on the intrinsic properties of the multi-particle Fock-states of the
vector-meson and some appear as meson-mass corrections to the
two-particle valence state. The latter ones, described in terms of
the same parameters as the leading-twist distribution amplitudes, turn
out to be numerically dominant; this is very welcome, as the matrix
elements describing the multi-particle states are only poorly known. To be
specific, putting all intrinsic higher-twist parameters $\zeta$ in
Table~\ref{tab:para2} to zero, the form factors change by at most
3\%. Hence, we conclude that the light-cone expansion is under good
control as well.

The dependence of form factors on the sum-rule parameters is small,
too. Changing $m_b$ by $\pm 100\,$MeV makes a 5\% effect at most and
is most pronounced at large $q^2$; at $q^2=0$ it is a 0.8\%
effect. This result means that, as for radiative corrections, there
is a strong cancellation of the $m_b$ dependence in the ratio of the
light-cone correlation function to $f_B$. The
same statement holds for the dependence on the continuum-threshold
within the limits specified in Table~\ref{tab:fB}. For the dependence
on the Borel parameter, we find an $\sim 7\%$ effect, increasing with
$q^2$, which again reminds us of the fact that the light-cone sum
rules become less reliable for large $q^2$.

The overall normalization of the form factors depends
on the vector-meson decay constants $f_V$ and $f_V^T$, the former 
determined experimentally, the latter calculated from QCD sum
rules (see\ Table~\ref{tab:para}). The corresponding uncertainty is 
at most 3\%.

Adding up all the errors in quadrature, we obtain an
uncertainty of the form factors of $\sim\,$11\%.

\begin{figure}
\centerline{\epsffile{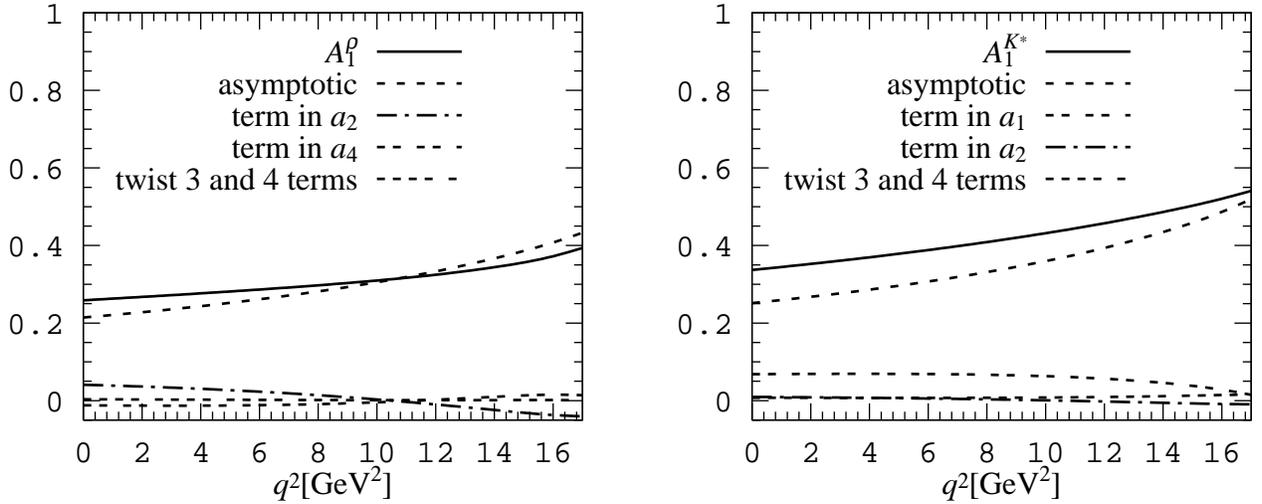}}
\caption[]{Separate contributions to the form factors $A_1^\rho$ 
and $A_1^{K^*}$.}\label{fig:X} 
\end{figure}

The shape of leading-twist distribution amplitudes, characterized
by the Gegenbauer
moments $a_{2,\rho}^{\parallel,\perp}$ for the $\rho$ and
$a_{1,K^*}^{\parallel,\perp}$, $a_{2,K^*}^{\parallel,\perp}$ for
the $K^*$, affects most significantly  the slope of the form factors
 and is illustrated in Fig.~\ref{fig:X} by two examples: $A_1^\rho$ and
$A_1^{K^*}$. The curves labelled ``asymptotic'' designate the form
factors as obtained by putting the $a_i$ to zero in
Eqs.~(\ref{eq:phipar}) and (\ref{eq:phiperp}); the corresponding
meson distribution amplitudes are completely model-independent and
dictated by perturbative QCD. The curves labelled
``$a_i$'' show corrections to this limit, which take into account
non-perturbative corrections to the distribution amplitudes; for illustration
we assumed in this figure the value $a_{4,\rho}^\perp =
a_{4,\rho}^\parallel = 0.1$ at $\mu = 1\,$GeV as a ball-park estimate 
for potential higher-order terms; this contribution is
not included in the final results. The curves labelled
``twist-3 and 4 terms'' show the contribution induced by the $\zeta$'s
in Table~\ref{tab:para2} and for the $K^*$ also contain terms
explicitly proportional to the strange quark-mass. It is obvious that
the ``asymptotic'' contribution grossly dominates, and
the remaining terms only add marginal corrections. It is also
obvious that the twist-3 and 4 terms do not have much overall
influence, whereas the contribution in $a_2$ (for $A_1^\rho$) and
$a_1$ (for $A_1^{K^*}$) tend to slow down the increase of the form
factors as functions of $q^2$. All involved parameters (except the
couplings $f_V$ and $f_V^T$) come with considerable theoretical
uncertainty. However, the only important error is that 
in $a_{2,\rho}$ and $a_{1,K^*}$:
it contributes $\sim\,$10\%  to the uncertainty in our predictions. 
Adding this number (in quadrature)
to the $\sim 11\%$ error from other sources, we 
end up with a total uncertainty of light-cone sum-rule predictions
of order $\sim 15\%$, which is our final error estimate.
 An improvement is to be expected if future lattice calculations achieve
an accuracy better than that quoted in Table~\ref{tab:para}. 

A few remarks are in order on the pattern of SU(3)-breaking. 
It enters our calculation at the following places:
\begin{itemize}
\item difference in decay constants: $f_{K^*}/f_\rho\approx
  f^T_{K^*}/f^T_\rho  = 1.14$,
  $f_{B_s}/f_B = 1.08$;
\item different meson-masses and continuum-thresholds $s_0$
  (Table~\ref{tab:fB});
\item different vector-meson distribution amplitudes
  (Table~\ref{tab:para}).
\end{itemize}
Figure~\ref{fig:X} also illustrates the relative size of these effects:
the difference between the ``asymptotic'' curves is almost exclusively
due to the
difference in $f_\rho$ and $f_{K^*}$ and makes a 17\% effect. For
$K^*$, the $a_2$ are small, whereas the $a_1$ are large and thus
increase the form factor. For $B_s\to\bar{K}^*$ decays, the sign
in $a_1$ is negative and $f_{B_s}$ is larger than $f_B$, so that we
observe considerably smaller form factors, see
Table~\ref{tab:fits}. The total SU(3)-breaking corrections
amount to $\sim\,$35\%, half of which comes from the decay constants
and half from the bigger momentum carried by the $s$ quark in the
strange hadron. Specifically, for $B_{u,d}$-decay form factors at
$q^2=0$ we obtain the values given in Table~\ref{tab:SU3}.

\begin{table}
$$
\renewcommand{\arraystretch}{1.5}
\addtolength{\arraycolsep}{3pt}
\begin{array}{|c|ccccc|}\hline
F & A_1 & A_2 & V & T_1 & T_3\\ \hline 
F^{K^*}(0)/F^\rho(0) & 1.30\pm 0.13 & 1.28\pm 0.13 & 1.36\pm 0.14 &
1.33\pm 0.13 & 1.29\pm 0.13\\ \hline
\end{array}
\renewcommand{\arraystretch}{1}
\addtolength{\arraycolsep}{-3pt}
$$
\caption[]{Size of SU(3)-breaking for $B_{u,d}$-decays into $\rho$ or
  $K^*$.}\label{tab:SU3}
\end{table}

\begin{figure}
\centerline{\epsffile{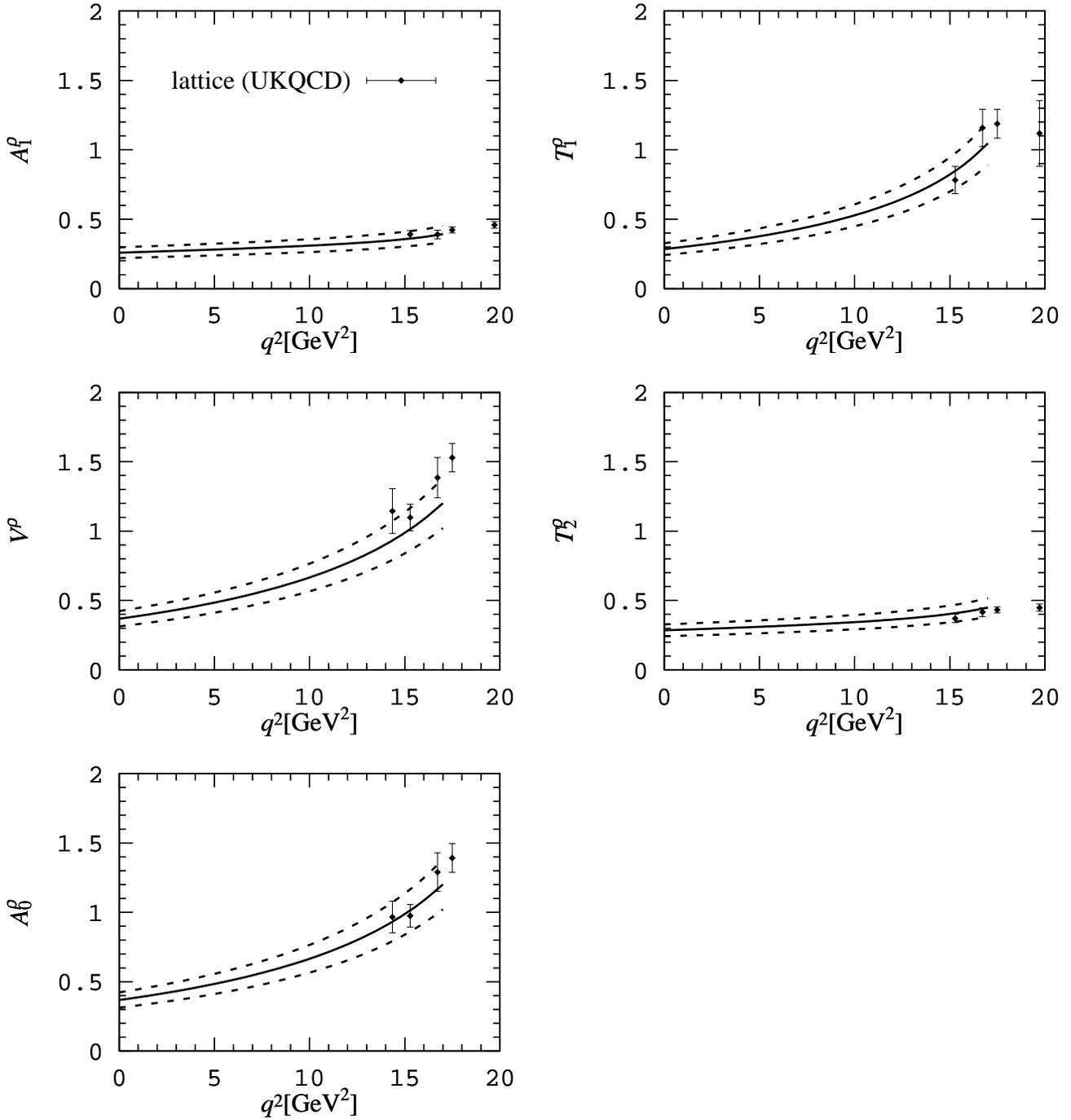}}
\caption[]{Comparison of the light-cone sum-rule predictions 
for the $B\to\rho$ form factors with lattice 
calculations \protect{\cite{UKQCD,latticeconstraints}}. Lattice errors 
are statistical only.
The dashed curves show the 15\% uncertainty range.
}\label{fig:lattrho}
\end{figure}

\begin{figure}
\centerline{\epsffile{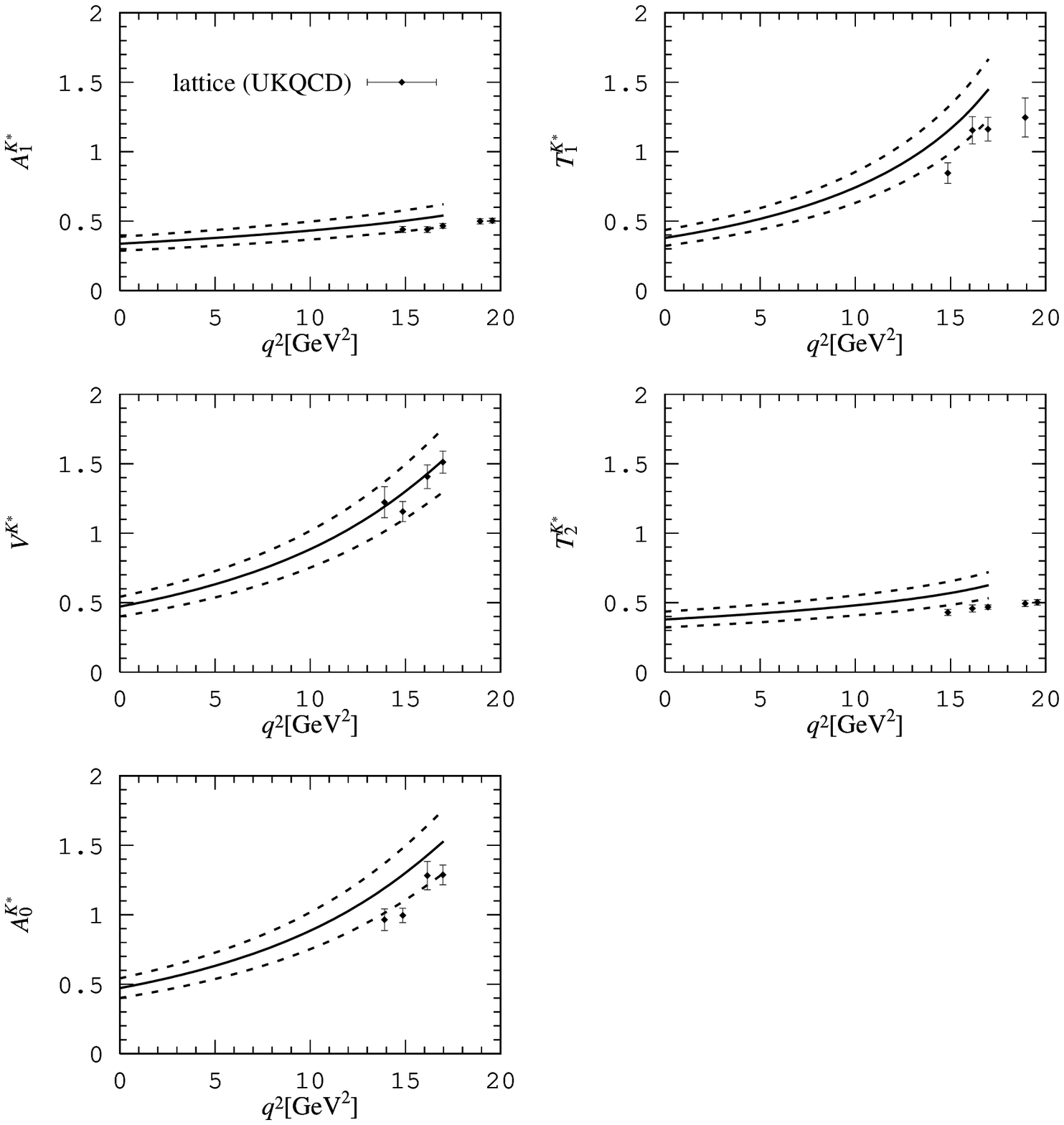}}
\caption[]{Comparison of the light-cone sum-rule predictions 
for the $B\to K^*$ form factors with lattice 
calculations \protect{\cite{UKQCD,latticeconstraints}}. Lattice 
errors are statistical only.
The dashed curves show the 15\% uncertainty range.
}\label{fig:lattKstar}
\end{figure}

In Fig.~\ref{fig:lattrho} we present a comparison of 
our results for $B\to\rho$ semileptonic and rare radiative form factors 
with the lattice calculations by the UKQCD collaboration 
\cite{UKQCD,latticeconstraints}. 
%
%
The agreement is very good. We wish to emphasize that the light-cone
sum-rule approach is theoretically more sound at small values of 
$q^2$, and in this sense is complementary to lattice techniques, which
work best  in the large $q^2$ region. 
A similar comparison for $B\to K^*$ decays is 
presented in Fig.~\ref{fig:lattKstar}. 
%
%
The agreement is somewhat worse
in this case; the lattice data favour smaller SU(3)-breaking effects.
This question deserves  further study.    

Finally, in Table~\ref{tab:comparison} we present
a comparison of the results of this work for the form factor values 
at $q^2=0 $ with earlier sum-rule calculations and the  
lattice results obtained using the light-cone sum-rule constraints.
\begin{table}
\renewcommand{\arraystretch}{1.5}
\addtolength{\arraycolsep}{3pt}
$$
\begin{array}{|l|lllll|}\hline
& {\rm This\ work} & \cite{ABS,rhoFFs} &
\cite{aliev}& \cite{latticeconstraints} {\rm
  (lattice} & \cite{3ptSR} \\[-7pt] 
 & {\rm (LCSR)} & {\rm(LCSR)} & {\rm(LCSR)} & {\rm +LCSR)} & 
{\rm (3pt SR)}\\ \hline
A_1^\rho(0)  & 0.26\pm0.04 & 0.27\pm0.05 & 0.30\pm0.05 &
0.27^{+0.05}_{-0.04} & 0.5\pm0.1\\
A_2^\rho(0)  & 0.22\pm0.03 & 0.28\pm0.05 & 0.33\pm0.05 &
0.26^{+0.05}_{-0.03} & 0.4\pm0.2\\
V(0)^\rho    & 0.34\pm0.05 & 0.35\pm0.07 & 0.37\pm0.07 &
0.35^{+0.06}_{-0.05} & 0.6\pm0.2\\
T_1^\rho(0)  & 0.29\pm0.04 & 0.24\pm0.07 & 0.30\pm0.10 & - & -\\
T_3^\rho(0)  & 0.20\pm0.03 & -          & 0.20\pm0.10 & - & -\\
A_1^{K^*}(0) & 0.34\pm0.05 & 0.32\pm0.06 & 0.36\pm0.05 &
0.29^{+0.04}_{-0.03} & 0.37\pm 0.03\\
A_2^{K^*}(0) & 0.28\pm0.04 & -          & 0.40\pm0.05 & - & 0.40\pm0.03\\
V^{K^*}(0)   & 0.46\pm0.07 & 0.38\pm0.08 & 0.45\pm0.08 & - & 0.47\pm0.03\\
T_1^{K^*}(0) & 0.38\pm0.06 & 0.32\pm0.05 & 0.34\pm0.10 &
0.32^{+0.04}_{-0.02} & 0.38\pm0.06\\
T_3^{K^*}(0) & 0.26\pm0.04 & -          & 0.26\pm0.10 & - & 0.6\\
\hline
\end{array}
$$
\renewcommand{\arraystretch}{1}
\addtolength{\arraycolsep}{-3pt}
\caption[]{Comparison of results from different works on form factors
  at $q^2=0$.}\label{tab:comparison}
\end{table}

\section{The Heavy-Quark Limit}
\setcounter{equation}{0}

The behaviour of $B$-decay form factors in the limit $m_b\to\infty$ is
interesting for various reasons. This limit was already discussed 
in some detail in Refs.~\cite{ABS,rhoFFs,BPi} so that in this paper
we only summarize the main points.

The first question concerns the scaling behaviour of form factors 
as functions of the $b$-quark mass. The behaviour depends on the 
momentum-transfer and is different for small and large recoil.
For $q^2\to0$, or, more precisely, for $m_b^2-q^2 \sim O(m_b^2)$, all 
form factors in question scale as $\sim 1/m_b^{3/2}$. This behaviour 
can be proved in perturbative QCD, taking into account 
Sudakov suppression of large transverse distances, but it is not restricted
to this regime and extends to ``soft'' terms as well \cite{rhoFFs,VMBreview}.  
For $m_b^2-q^2\sim O(m_b)$, on the other hand, the form factors 
obtained from light-cone  sum-rules satisfy the scaling behaviour predicted 
by Heavy-Quark Effective Theory (HQET) \cite{IW90}.
For realistic values of the $b$-quark mass, these two regimes are not well 
separated; therefore large corrections to asymptotic 
scaling are to be expected. Some estimates of pre-asymptotic corrections
are presented in Refs.~\cite{ABS,rhoFFs}. They have to be considered
as indicative only. We do not attempt to further quantify 
pre-asymptotic corrections in this work.  

\begin{figure}
\centerline{\epsffile{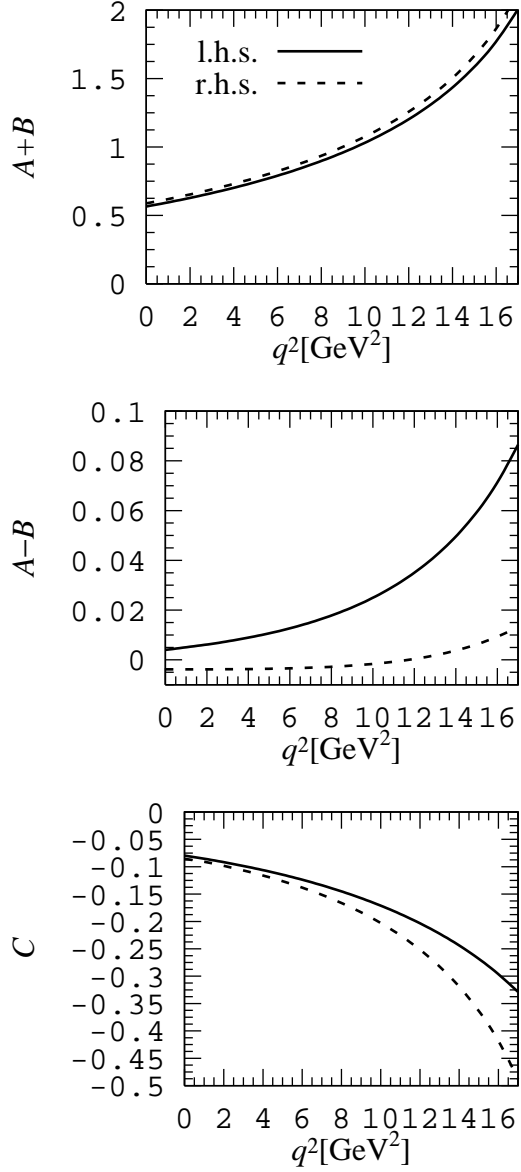}}
\caption[]{Isgur-Wise relations 
(\protect{\ref{eq:IW4}})--(\protect{\ref{eq:IW6}}) for $B\to\rho$
transitions. Renormalization scale is $\mu=m_b$.
Solid and dashed curves correspond to expressions appearing on the l.h.s.\ and
r.h.s., respectively.}\label{fig:IW2}
\end{figure}

The second question concerns possible relations between different 
form factors in the heavy-quark limit.  Heavy-quark symmetry implies 
exact relations between semileptonic and penguin 
form factors at small recoil and renormalization scale $\mu=m_b$ 
\cite{IW90}, which, using the penguin form factor definitions in 
Eq.~(\ref{eq:ABC}), can conveniently be 
written as:
\begin{eqnarray}
  A(q^2)+B(q^2) &=& \frac{2 m_B}{m_B+m_V}\,V(q^2),
\label{eq:IW4}\\
  A(q^2)-B(q^2) &=& \frac{(m_B+m_V)}{m_B}\, A_1(q^2) - 
         \frac{m_B^2-q^2+m_V^2}{m_B}\frac{V(q^2)}{m_B+m_V},
\label{eq:IW5}\\
  C(q^2) &=& - \frac{m_B-m_V}{m_B}\, V(q^2) 
+\frac{m_V(m_B^2-m_V^2)}{m_B q^2} \left[A_0(q^2)-A_3(q^2)\right]
\nonumber\\
&&{}+\frac{m_B-m_V}{2m_B}\, A_2(q^2).
\label{eq:IW6}
\end{eqnarray} 
Writing the relations in this form emphasizes their different behaviour
in the heavy-quark limit: at small recoil, both sides of
Eqs.~(\ref{eq:IW4}) and  (\ref{eq:IW6}) are of order $\sqrt{m_b}$, 
while Eq.~(\ref{eq:IW5}) relates combinations of form factors, which are 
of order $1/\sqrt{m_b}$. The numerical comparison for $B\to\rho$
transitions is presented in 
Fig.~\ref{fig:IW2}.
We note that (a) Eq.~(\ref{eq:IW4}) is satisfied with high accuracy. 
(b) The relation (\ref{eq:IW5}) is violated. However, both sides 
are numerically small compared to Eq.~(\ref{eq:IW4}), in agreement with
the expected $1/m_b$ suppression. (c) The relation (\ref{eq:IW6})
is  very well satisfied at $q^2\to0$ and it holds with 20\% accuracy 
at large $q^2$; both sides turn out to be small at large recoil,
which implies significant cancellations between the terms on the
right-hand side. 
%
%
%
\begin{figure}
\centerline{\epsffile{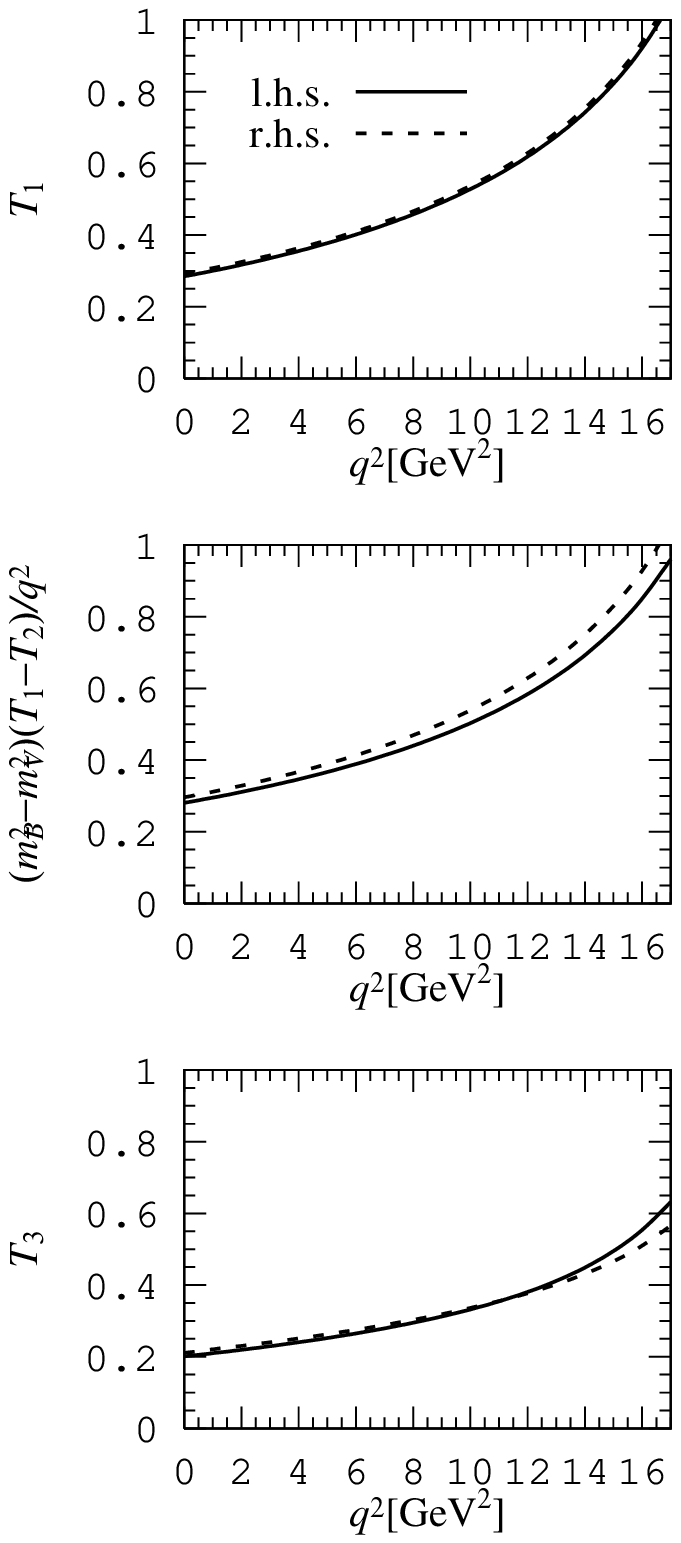}}
\caption[]{Isgur-Wise relations 
(\protect{\ref{eq:IW1}})--(\protect{\ref{eq:IW3}}) for $B\to\rho$
transitions. Renormalization scale is $\mu=m_b$.
Solid and dashed curves correspond to expressions appearing on the l.h.s.\ and
r.h.s., respectively.}\label{fig:IW}
\end{figure}
%

For phenomenological applications it is more appropriate to rewrite 
the Isgur-Wise relations  (\ref{eq:IW4})--(\ref{eq:IW6}) in terms of 
the form factors defined in (\ref{eq:T}):
\begin{eqnarray}
  T_1(q^2) &=& \frac{m_B^2+q^2-m_V^2}{2m_B}\frac{V(q^2)}
   {m_B+m_V}+\frac{m_B+m_V}{2m_B} A_1(q^2),
\label{eq:IW1}\\
\hspace*{-0.5cm}\frac{m_B^2-m_V^2}{q^2}\Big[T_1(q^2)-T_2(q^2)\Big] &=&
   \frac{3m_B^2-q^2+m_V^2}{2m_B}\frac{V(q^2)}
   {m_B+m_V}-\frac{m_B+m_V}{2m_B} A_1(q^2),
\label{eq:IW2}\\
  T_3(q^2) &=& \frac{m_B^2-q^2+3m_V^2}{2m_B} \frac{V(q^2)}{m_B+m_V}+
\frac{m_B^2-m_V^2}{m_B q^2} m_V A_0(q^2) 
\nonumber\\  &\hspace*{-1cm}-&\hspace*{-0.5cm}\frac{m_B^2+q^2-m_V^2}{2m_B q^2}
  \Big[(m_B+m_V)A_1(q^2)-(m_B-m_V)A_2(q^2)\Big].
\label{eq:IW3} 
\end{eqnarray}
Note that such a rewriting mixes terms of different order in $1/m_b$
in the small recoil region, and in this sense is not fully consistent
with the derivation in \cite{IW90}.
It can be justified, however, by observing that the hierarchy of 
contributions is different at large recoil and all the terms
become formally of the same order. The numerical comparison for
$B\to\rho$ transitions is 
presented in Fig.~\ref{fig:IW}.
The  accuracy proves to be excellent for the relation (\ref{eq:IW1}),
which is observed to within 3\% accuracy, and good for (\ref{eq:IW2})
with deviations of at most 8\%. Relation (\ref{eq:IW3}), however, is
violated by 20\% for $q^2>15\,$GeV$^2$.
Since fidelity of the sum-rules worsens in the high-$q^2$ region, 
it is not clear whether this disagreement indicates a genuine $1/m_b$ 
correction or is an artefact. 
Our results reinforce an earlier observation in
\cite{ABS} that the relation in (\ref{eq:IW1}) is satisfied within 
$\sim\,$(5--7)\% in the whole region of $q^2$
to  leading-twist accuracy in the light-cone sum-rule approach, 
and strongly support the conjecture of \cite{BD91} about the  
validity of heavy-quark symmetry relations in the region of small $q^2$
in heavy-to-light transitions.

\section{Conclusions}
We have given a complete analysis of $B$-decay form factors to light 
vector-mesons in the light-cone sum-rule approach. The principal
new contribution of this work is the calculation of 
radiative corrections and higher-twist
corrections to the sum-rules, which are calculated for the first time.
We observe that the light-cone sum-rules turn out to be very robust 
against corrections in the light-cone expansion, whose numerical 
impact proves to be minimal. Radiative corrections seem to be well
under control. In  cases where higher-twist corrections
are important, they are dominated by meson-mass effects, which do not 
involve free parameters. 
The theoretical accuracy of the approach is thus restricted entirely by the  
duality approximation for the extraction of the $B$-meson state from the 
continuum and contributions of higher resonances. The usual ``educated guess''
is that the accuracy of such an extraction is of order 10\%, which provides
an irreducible error. Effects of yet higher radiative 
corrections and yet higher-twists are likely to be much less; therefore,
the sum-rules derived in this work cannot be improved significantly.
The numerical analysis, however, can and should eventually be updated,
once estimates for the meson 
distribution amplitudes, $b$-quark mass and $f_B$ become more precise. 
In particular, lattice calculations of the tensor 
couplings $f^T_V$ and the parameters $a_{1,2}^\parallel$, 
$a_{1,2}^\perp$  for meson
distribution amplitudes would be most welcome.
  
\section*{Acknowledgements}
P.B. is grateful to NORDITA for hospitality and partial support
during her visit when part of this work was done. We thank E.~Bagan
for collaboration in early stages of this work and L.~Lellouch for 
providing us with the new data by the UKQCD collaboration. 
 
\setcounter{equation}{0}


\appendix
\renewcommand{\theequation}{\Alph{section}.\arabic{equation}}
\setcounter{table}{0}
\renewcommand{\thetable}{\Alph{table}}

\section{One-loop Integrals}

For the calculation of radiative corrections, we need the following
integrals:
\begin{eqnarray}
\int{d^Dk \over(2\pi)^D} {k_\alpha \over (k+up)^2 k^2 [(q-k)^2-m^2]}
&=& A up_{\alpha}+Bq_\alpha, \\
\int{d^Dk \over(2\pi)^D} {k_\alpha (q-k)_\beta \over (k+up)^2 k^2 
[(q-k)^2-m^2]} &=& C g_{\alpha\beta}+
D q_\alpha q_\beta
+E q_\alpha up_{\beta}+F
up_{\alpha} q_\beta+\dots,\\
\lefteqn{\int{d^Dk \over(2\pi)^D} \frac{k_\alpha  k_\beta}{ k^2  (k-u
    p)^2 (k+\bar u p)^2
[(u p+q-k)^2-m^2]}\ =\ H g_{\alpha\beta}+I q_\alpha
q_\beta +\dots,}\hspace*{6.5cm}\\
\int{d^Dk \over(2\pi)^D} {1 \over k^2 [(up+q-k)^2-m^2]} &=& Y, \\
\int{d^Dk \over(2\pi)^D} {k_\alpha\over k^2 [(up+q-k)^2-m^2]}
&=&Z\;(q+up)_\alpha,
\end{eqnarray}
where the dots stand for terms that are irrelevant to the present 
calculation. The functions
$\bar A$, $\bar B$, $\bar C$ are obtained from $A$, $B$ and $C$
by the replacement
\begin{equation}
u\to \bar u;\qquad q\to -p_B.
\end{equation}
We shall use the notation
\begin{equation}
s\equiv m^2-u p_B^2-\bar u q^2,\quad \frac{1}{\hat\epsilon} =
\frac{1}{\epsilon} -\gamma_E + \ln\, 4\pi,
\end{equation}
with $D= 4-2\epsilon$.
In order to perform Borel transformation and continuum subtraction, the 
following spectral representations for the above integrals prove  
the most convenient:
\begin{eqnarray}
A & = & {i \over(4\pi)^2}
\int\limits_{m^2}^\infty {{\rm d}t\over t-\xi}{1\over (t-q^2)^2}
\left\{
(m^2-q^2)\left[
-{1\over\hat{\epsilon}}-1+\log{(t-m^2)^2\over\mu^2 t}
\right]+t-q^2-{q^2(m^2-t)\over t}
\right\}\nonumber\\
\lefteqn{u (p_B^2-q^2) A \ =\ {{\rm i}\over(4\pi)^2}\left\{
{1\over\hat\epsilon}+2-\log{s\over\mu^2}
+\int_{m^2}^\infty
{{\rm d}t\over t-\xi}
\right.}\nonumber
\\
& & \hspace*{1.9cm}\times\left.{1\over t-q^2}\left[
(m^2-q^2)\left(
-{1\over\hat\epsilon}-1+\log{(m^2-t)^2\over\mu^2 t}
\right)-{q^2\over t}(m^2-t)
\right]
\phantom{\int_{m_b^2}^\infty}\kern-0.7cm
\right\}\nonumber\\
\lefteqn{\bar u (q^2-p_B^2) \bar A \ =\ 
{{\rm i}\over(4\pi)^2}\left\{
{1\over\hat\epsilon}+2-\log{s\over\mu^2}
+\int_{m^2}^\infty
{{\rm d}t\over t-\xi}
\right.}\nonumber
\\
& &\hspace*{1.5cm}\times\left.\left[
\left(1+{m^2-t\over  t-p^2_B}\right)\left(
-{1\over\hat\epsilon}-1+\log{(m^2-t)^2\over\mu^2 t}
\right)+\left({1\over t}-{1\over t-p_B^2}\right)(m^2-t))
\right]
\phantom{\int_{m^2}^\infty}\kern-0.7cm
\right\}\nonumber\\
B & = & {i\over(4\pi)^2}\int_{m^2}^\infty {{\rm d}t\over t-\xi}
{m^2-t\over t(t-q^2)}\nonumber\\
\bar B&=&{i \over(4\pi)^2}{1\over\bar u}
\int_{m^2}^\infty {{\rm d} t\;(m^2-t)\over t(t-q^2)}
\left(
{1\over t-p_B^2}-{u\over t-\xi}
\right)\nonumber\\
C & = & {i \over(4\pi)^2}{1\over4}\left\{
-{1\over\hat{\epsilon}}-3+\log{m^2-\xi\over\mu^2}+
\int_{m^2}^\infty {{\rm d}t\over t-\xi}
{(2m^2-q^2)t-m^4\over t(t-q^2)}
\right\}\nonumber\\
\bar C&=&{i \over(4\pi)^2}{1\over4}
\left\{
-{1\over\hat{\epsilon}}-3+\log{s\over\mu^2}+\int_{m^2}^\infty {\rm d} t\;\left[
{1\over t-\xi}-{1\over\bar u}{(m^2-t)^2\over  t(t-q^2)}
\left(
{1\over t-p_B^2}-{u\over t-\xi}
\right)
\right]
\right\}\nonumber\\
D & = & {i\over(4\pi)^2}{1\over2}\int_{m^2}^\infty {{\rm d}t\over t-\xi}
{m^4-t^2\over t^2(t-q^2)}\nonumber\\
E & = & \frac{i}{(4\pi)^2}\,\frac{1}{2u(p_B^2-q^2)} +
\frac{1}{u(p_B^2-q^2)}\,(m^2 B - q^2 D)\nonumber\\
F & = & A+E\nonumber\\
H&=&{i\over(4\pi)^2}{1\over2}\int_{m^2}^\infty {\rm d} t\;
\left\{
{1\over t-\xi}\left[{1\over\hat{\epsilon}}+2-\log{(t-m^2)^2\over t\mu^2}\right]
\left[{u(m^2-q^2)\over(t-q^2)^2}+
{\bar u(m^2-t)\over(t-p_B^2)^2}+{\bar u \over t-p_B^2}\right]\right.\nonumber\\
&&\phantom{{{\rm
      i}\over(4\pi)^2}{1\over2}\int_{m^2}^\infty 
{\rm d}
t\;} -\left.{1\over(t-p_B^2)(t-q^2)}{t+m^2\over t}
\right\}\nonumber\\
I & = & {i\over(4\pi)^2}\int_{m^2}^\infty {\rm d} t\;
\left\{
{m^2-t\over(t-\xi)t}\left[
{u\over(t-q^2)^2}+{\bar u\over(t-p_B^2)^2}
\right]+{m^2\over t^2(t-p_B^2)(t-q^2)}
\right\}\nonumber\\
Y & = & \frac{i}{(4\pi)^2}\left(\frac{1}{\hat\epsilon} - 
\ln\,\frac{s}{\mu^2} + 2
-\frac{m^2}{m^2-s}\,\ln \frac{m^2}{s}\right)\nonumber\\
Z & = & \frac{1}{2}\,\frac{i}{(4\pi)^2}\left( \frac{1}{\hat\epsilon} -
\ln\,\frac{m^2}{\mu^2} - \frac{m^2}{m^2-s} + 2 -
\frac{s^2}{(m^2-s)^2}\,\ln\, \frac{s}{m^2}\right).
\end{eqnarray}

\section{Summary of Meson Distribution Amplitudes}\label{app:B}
\setcounter{equation}{0}

The expressions collected in this appendix are principally the result
of recent studies reported in Refs.~\cite{BBrho,BBKT,BBS}. 
We use a simplified version of the set of twist-4 distributions \cite{BBS},
taking into account only contributions of the lowest conformal partial-waves,
and for consistency discard contributions of higher partial-waves in 
twist-3 distributions in cases where they enter physical amplitudes 
multiplied by additional powers of $m_\rho$. 
The SU(3)-breaking effects are taken 
into account in leading-twist distributions and partially 
for twist-3, but neglected for twist-4. Explicit expressions are given 
below for a (charged) $\rho$-meson. 
Distribution amplitudes for other vector-mesons are obtained 
by trivial substitutions.

Throughout this appendix we denote the meson momentum by $P_\mu$  
and introduce the light-like vectors $p$ and $z$ such that 
\begin{equation}
p_\mu = P_\mu-\frac{1}{2}z_\mu \frac{m^2_\rho}{pz}.
\label{smallp}
\end{equation}  
The meson polarization vector $e^{(\lambda)}_\mu$ is decomposed in 
projections onto the two light-like vectors and the orthogonal plane 
as
\begin{equation}
 e^{(\lambda)}_\mu = \frac{(e^{(\lambda)}\cdot z)}{pz}
\left( p_\mu -\frac{m^2_\rho}{2pz} z_\mu \right)+e^{(\lambda)}_{\perp\mu}. 
\label{polv}
\end{equation} 

\subsection{Chiral-even distributions}
Two-particle quark-antiquark distribution amplitudes are defined 
as matrix elements of non-local operators on the light-cone \cite{BBKT}:
\begin{eqnarray}
\langle 0|\bar u(z) \gamma_{\mu} d(-z)|\rho^-(P,\lambda)\rangle 
&=& f_{\rho} m_{\rho} \left[ p_{\mu}
\frac{e^{(\lambda)}\cdot z}{p \cdot z}
\int_{0}^{1} \!du\, e^{i \xi p \cdot z} \phi_{\parallel}(u, \mu^{2}) \right. 
+ e^{(\lambda)}_{\perp \mu}
\int_{0}^{1} \!du\, e^{i \xi p \cdot z} g_{\perp}^{(v)}(u, \mu^{2}) 
\nonumber \\
& & \left.{}- \frac{1}{2}z_{\mu}
\frac{e^{(\lambda)}\cdot z }{(p \cdot z)^{2}} m_{\rho}^{2}
\int_{0}^{1} \!du\, e^{i \xi p \cdot z} g_{3}(u, \mu^{2}) \right]
\label{eq:vda}
\end{eqnarray}
and 
\begin{equation}
\langle 0|\bar u(z) \gamma_{\mu} \gamma_{5} 
d(-z)|\rho^-(P,\lambda)\rangle = 
 \frac{1}{2}\left(f_{\rho} - f_{\rho}^{T}
\frac{m_{u} + m_{d}}{m_{\rho}}\right)
m_{\rho} \epsilon_{\mu}^{\phantom{\mu}\nu \alpha \beta}
e^{(\lambda)}_{\perp \nu} p_{\alpha} z_{\beta}
\int_{0}^{1} \!du\, e^{i \xi p \cdot z} g^{(a)}_{\perp}(u, \mu^{2}).
\label{eq:avda}
\end{equation}
 For brevity, here and below we do not show the gauge factors between the 
quark and the antiquark fields and use the short--hand notation
$$\xi = u - (1-u) = 2u-1.$$ 
The vector and tensor  decay constants $f_\rho$ and $f_\rho^T$ are defined,
as usual, as
\begin{eqnarray}
\langle 0|\bar u(0) \gamma_{\mu}
d(0)|\rho^-(P,\lambda)\rangle & = & f_{\rho}m_{\rho}
e^{(\lambda)}_{\mu},
\label{eq:fr}\\
\langle 0|\bar u(0) \sigma_{\mu \nu} 
d(0)|\rho^-(P,\lambda)\rangle &=& i f_{\rho}^{T}
(e_{\mu}^{(\lambda)}P_{\nu} - e_{\nu}^{(\lambda)}P_{\mu}).
\label{eq:frp}
\end{eqnarray}
The distribution amplitude $\phi_\parallel$ is of twist-2,
$g_\perp^{(v)}$ and $g_\perp^{(a)}$ are twist-3 and $g_3$ is twist-4. 
All four functions $\phi=\{\phi_\parallel,
g_\perp^{(v)},g_\perp^{(a)},g_3\}$ are normalized as
\begin{equation}
\int_0^1\!du\, \phi(u) =1,
\label{eq:norm}
\end{equation}
which can be checked by comparing the two sides
of the  defining equations in the limit $z_\mu\to 0$ and using the
equations of motion.
We keep the (tiny) corrections proportional 
to the $u$ and $d$ quark-masses $m_u$ and $m_d$ to indicate the 
SU(3)-breaking corrections for $K^*$- and $\phi$-mesons.

In addition, we have to define three-particle distributions:
\begin{eqnarray}
\langle 0|\bar u(z) g\widetilde G_{\mu\nu}\gamma_\alpha\gamma_5 
  d(-z)|\rho^-(P,\lambda)\rangle &=&
  f_\rho m_\rho p_\alpha[p_\nu e^{(\lambda)}_{\perp\mu}
 -p_\mu e^{(\lambda)}_{\perp\nu}]{\cal A}(v,pz)
\nonumber\\ &&{}
+ f_\rho m_\rho^3\frac{e^{(\lambda)}\cdot z}{pz}
[p_\mu g^\perp_{\alpha\nu}-p_\nu g^\perp_{\alpha\mu}] \widetilde\Phi(v,pz)
\nonumber\\&&{}
+ f_\rho m_\rho^3\frac{e^{(\lambda)}\cdot z}{(pz)^2}
p_\alpha [p_\mu z_\nu - p_\nu z_\mu] \widetilde\Psi(v,pz)
\end{eqnarray}
\begin{eqnarray}
\langle 0|\bar u(z) g G_{\mu\nu}i\gamma_\alpha 
  d(-z)|\rho^-(P)\rangle &=&
  f_\rho m_\rho p_\alpha[p_\nu e^{(\lambda)}_{\perp\mu} 
  - p_\mu e^{(\lambda)}_{\perp\nu}{\cal V}(v,pz)
\nonumber\\&&{}
+ f_\rho m_\rho^3\frac{e^{(\lambda)}\cdot z}{pz}
[p_\mu g^\perp_{\alpha\nu} - p_\nu g^\perp_{\alpha\mu}] \Phi(v,pz)
\nonumber\\&&{}
+ f_\rho m_\rho^3\frac{e^{(\lambda)}\cdot z}{(pz)^2}
p_\alpha [p_\mu z_\nu - p_\nu z_\mu] \Psi(v,pz),
\end{eqnarray}
where 
\begin{equation}
   {\cal A}(v,pz) =\int {\cal D}\underline{\alpha} 
e^{-ipz(\alpha_u-\alpha_d+v\alpha_g)}{\cal A}(\underline{\alpha}),
\end{equation}
etc., and $\underline{\alpha}$ is the set of three momentum fractions
$\underline{\alpha}=\{\alpha_d,\alpha_u,\alpha_g\}$.
 The integration measure is defined as 
\begin{equation}
 \int {\cal D}\underline{\alpha} \equiv \int_0^1 d\alpha_d
  \int_0^1 d\alpha_u\int_0^1 d\alpha_g \,\delta(1-\sum \alpha_i).
\label{eq:measure}
\end{equation}
The distribution amplitudes ${\cal V}$ and ${\cal A}$ are of twist-3,
while the rest is twist-4 and we have not shown further Lorentz structures 
corresponding to twist-5 contributions\footnote{We use  
a different normalization of three-particle twist-3 distributions
compared to \cite{BBKT}.}.

Calculation of exclusive amplitudes involving a large momentum-transfer
reduces to evaluation of meson-to-vacuum transition matrix elements
of non-local operators, which can be expanded in powers of the deviation
from the light-cone (see text).  
To twist-4 accuracy one can use the expression for the axial-vector matrix 
element in (\ref{eq:avda}) as it stands, replacing the light-cone 
vector $z_\mu$ by the actual quark-antiquark separation $x_\mu$.
For the vector operator, the light-cone expansion to the twist-4 accuracy 
reads:
\begin{eqnarray}
\langle0|\bar u(x) \gamma_\mu d(-x)|\rho^-(P,\lambda)\rangle 
 &=&
f_\rho m_\rho \Bigg\{
\frac{e^{(\lambda)}x}{Px}\int_0^1 du \,e^{i\xi Px}
\Big[\phi_\parallel(u,\mu)
+\frac{m^2_\rho x^2}{4}  {\Bbb A}(u,\mu)\Big]
\nonumber\\
&&{}+\left(e^{(\lambda)}_\mu-P_\mu\frac{e^{(\lambda)}x}{Px}\right)
\int_0^1 du\, e^{i\xi Px} \,g^{(v)}_\perp(u,\mu)
\nonumber\\
&&{}-\frac{1}{2}x_\mu \frac{e^{(\lambda)}x}{(Px)^2} m^2_\rho \int_0^1 du 
\, e^{i\xi Px} {\Bbb C} (u,\mu)
\Bigg\},
\label{eq:OPE1}
\end{eqnarray}
where
\begin{equation}
  {\Bbb C}(u) = g_3(u)+\phi_\parallel(u) -2 g^{(v)}_\perp(u)
\end{equation}
and ${\Bbb A}(u)$ can be related to integrals of three-particle distributions 
using equations of motion. All distribution functions in (\ref{eq:OPE1})
are assumed to be normalized at the scale $\mu^2\sim x^{-2}$ (to 
leading-logarithmic accuracy). In practical calculations it is sometimes
convenient to use integrated distributions
\begin{equation}
 {\Bbb C}^{(i)}(u) = -\int_0^u\!dv\,{\Bbb C}(v),\qquad 
 {\Bbb C}^{(ii)}(u) = -\int_0^u\!dv\,{\Bbb C}^{(i)}(v).
\label{eq:c12}
\end{equation}

For the leading twist-2 distribution amplitude $\phi_\parallel$ we use
\begin{equation}\label{eq:phipar}
\phi_\parallel(u) =  6 u\bar u \left[ 1 + 3 a_1^\parallel\, \xi +
a_2^\parallel\, \frac{3}{2} ( 5\xi^2  - 1 ) \right]
\end{equation}
with parameter values as specified in Table~\ref{tab:para}.
The expressions for higher-twist distributions given below correspond
to the simplest self-consistent approximation that satisfies the  
QCD equations of motion \cite{BBKT,BBS}:
\begin{itemize}
\item{} Three-particle distributions of twist-3:
\begin{eqnarray}
{\cal V} (\underline{\alpha}) &=& 
540\, \zeta_3 \omega^V_3 (\alpha_d-\alpha_u)\alpha_d \alpha_u\alpha_g^2,
\label{modelV}\\
{\cal A} (\underline{\alpha}) &=& 
360\,\zeta_3 \alpha_d \alpha_u \alpha_g^2 
\Big[ 1+ \omega^A_{3}\frac{1}{2}(7\alpha_g-3)].
\label{modelA}
 \end{eqnarray}
\item{} Two-particle distributions of twist-3:
\begin{eqnarray}
g_\perp^{(a)}(u) & = & 6 u \bar u \left[ 1 + a_1^\parallel \xi +
\left\{\frac{1}{4}a_2^\parallel +
\frac{5}{3}\, \zeta_{3} \left(1-\frac{3}{16}\,
\omega^A_{3}+\frac{9}{16}\omega^V_3\right)\right\}
(5\xi^2-1)\right]\nonumber\\
& & {} + 6\, \widetilde{\delta}_+ \,  (3u \bar u + \bar u \ln \bar u +
u \ln u ) + 
6\, \widetilde{\delta}_- \,  (\bar u \ln \bar u - u \ln u),\\
 g_\perp^{(v)}(u) & = & \frac{3}{4}(1+\xi^2)
+ a_1^\parallel\,\frac{3}{2}\, \xi^3 
 + \left(\frac{3}{7} \, 
a_2^\parallel + 5 \zeta_{3} \right) \left(3\xi^2-1\right)
 \nonumber\\
& & {}+ \left[ \frac{9}{112}\, a_2^\parallel 
+ \frac{15}{64}\, \zeta_{3}\Big(3\,\omega_{3}^V-\omega_{3}^A\Big)
 \right] \left( 3 - 30 \xi^2 + 35\xi^4\right)\nonumber\\
& & {}+\frac{3}{2}\,\widetilde{\delta}_+\,(2+\ln u + \ln\bar u) +
\frac{3}{2}\,\widetilde{\delta}_-\, ( 2 \xi + \ln\bar u - \ln u),\label{eq:gv}
\end{eqnarray}
\item{} Three-particle distributions of twist-4:
\begin{eqnarray}
\widetilde \Phi (\underline{\alpha}) &=& 
  \Big[-\frac{1}{3}\zeta_{3}+\frac{1}{3}\zeta_{4}\Big] 
   30(1-\alpha_g)\alpha_g^2,
\nonumber\\
  \Phi (\underline{\alpha}) &=& 
  \Big[-\frac{1}{3}\zeta_{3}+\frac{1}{3}\zeta_{4}\Big] 
   30(\alpha_u-\alpha_d)\alpha_g^2,
\nonumber\\
  \widetilde\Psi (\underline{\alpha}) &=& 
  \Big[\frac{2}{3}\zeta_{3}+\frac{1}{3}\zeta_{4}\Big] 
   120 \alpha_u\alpha_d\alpha_g,
\nonumber\\
 \Psi (\underline{\alpha}) &=& 0.
\end{eqnarray}
\item{} Two-particle distributions of twist-4: 
\begin{eqnarray}
{\Bbb A}(u) &=& \Bigg[\frac{4}{5}+\frac{20}{9} \zeta_{4}
                    +\frac{8}{9} \zeta_{3}\Bigg]30 u^2(1-u)^2,
\nonumber\\
g_3(u) &=& 6u(1-u) + \Bigg[\frac{10}{3} \zeta_{4}
                    -\frac{20}{3} \zeta_{3}\Bigg](1-3 \xi^2),
\nonumber\\
{\Bbb C}(u) &=& \Bigg[\frac{3}{2}+\frac{10}{3} \zeta_{4}
                    +\frac{10}{3} \zeta_{3}\Bigg](1-3 \xi^2),
\nonumber\\
 {\Bbb C}^{(ii)}(u) &=& \Bigg[\frac{3}{2}+\frac{10}{3} \zeta_{4}
                    +\frac{10}{3} \zeta_{3}\Bigg]u^2(1-u)^2,
\end{eqnarray}
\end{itemize} 
where the dimensionless couplings $\zeta_3$ and $\zeta_4$ are 
defined as local matrix elements
\begin{eqnarray}
\langle0|\bar u g\tilde G_{\mu\nu}\gamma_\alpha
 \gamma_5 d|\rho^-(P,\lambda)\rangle  &=&
f_\rho m_\rho \zeta_{3}
\Bigg[
e^{(\lambda)}_\mu\Big(P_\alpha P_\nu-\frac{1}{3}m^2_\rho \,g_{\alpha\nu}\Big)  
-e^{(\lambda)}_\nu\Big(P_\alpha P_\mu-\frac{1}{3}m^2_\rho \,g_{\alpha\mu}\Big)
\Bigg]
\nonumber\\
&&{}+\frac{1}{3}f_\rho m_\rho^3 \zeta_{4}
\Bigg[e^{(\lambda)}_\mu g_{\alpha\nu}- e^{(\lambda)}_\nu g_{\alpha\mu}\Bigg]  
\end{eqnarray}
and have been estimated from QCD sum-rules 
 \cite{ZZC85,BK86}.
The terms in $\delta_\pm$ and $\widetilde\delta_\pm$ specify quark-mass 
corrections in twist-3 distributions induced by the equations of motion.
The numerical values of these and other coefficients are listed in 
Tables~\ref{tab:para} and \ref{tab:para2}\footnote{In the 
notation of Ref.~\cite{BBKT},
$\omega_{1,0}^A\equiv \omega_3^A$, $ \zeta_3^A\equiv \zeta_3$, and 
$\zeta_3^V \equiv (3/28)\zeta_3\omega_3^V$.}.
Note that we neglect SU(3)-breaking effects in twist-4 distributions
and in gluonic parts of twist-3 distributions.

\subsection{Chiral-odd distributions}

There exist four different two-particle chiral-odd distributions \cite{BBKT}
defined as
\begin{eqnarray}
\langle 0|\bar u(z) \sigma_{\mu \nu}  
d(-z)|\rho^-(P,\lambda)\rangle 
&=& i f_{\rho}^{T} \left[ ( e^{(\lambda)}_{\perp \mu}p_\nu -
e^{(\lambda)}_{\perp \nu}p_\mu )
\int_{0}^{1} \!du\, e^{i \xi p \cdot z} \phi_{\perp}(u, \mu^{2}) \right. 
\nonumber \\
& &{}+ (p_\mu z_\nu - p_\nu z_\mu )
\frac{e^{(\lambda)} \cdot z}{(p \cdot z)^{2}}
m_{\rho}^{2} 
\int_{0}^{1} \!du\, e^{i \xi p \cdot z} h_\parallel^{(t)} (u, \mu^{2}) 
\nonumber \\
& & \left.{}+ \frac{1}{2}
(e^{(\lambda)}_{\perp \mu} z_\nu -e^{(\lambda)}_{\perp \nu} z_\mu) 
\frac{m_{\rho}^{2}}{p \cdot z} 
\int_{0}^{1} \!du\, e^{i \xi p \cdot z} h_{3}(u, \mu^{2}) \right],
\label{eq:tda}
\end{eqnarray}
\begin{equation}
\langle 0|\bar u(z)
d(-z)|\rho^-(P,\lambda)\rangle
= -i \left(f_{\rho}^{T} - f_{\rho}\frac{m_{u} + m_{d}}{m_{\rho}}
\right)(e^{(\lambda)}\cdot z) m_{\rho}^{2}
\int_{0}^{1} \!du\, e^{i \xi p \cdot z} h_\parallel^{(s)}(u, \mu^{2}).
\label{eq:sda}
\end{equation}
The distribution amplitude $\phi_\perp$ is twist-2, $h_\parallel^{(s)}$ and $h_\parallel^{(t)}$ are
twist-3 and $h_3$ is twist-4.   
All four functions $\phi=\{\phi_\perp,h_\parallel^{(s)},h_\parallel^{(t)},h_3\}$ are normalized to
$\int_0^1\!du\, \phi(u) =1$. 

Three-particle chiral-odd distributions are defined to twist-4 accuracy as
\begin{eqnarray}
\lefteqn{\langle 0|\bar u(z) \sigma_{\alpha\beta}
         gG_{\mu\nu}(vz) 
         d(-z)|\rho^-(P,\lambda)\rangle\ =}\hspace*{2cm} 
 \nonumber \\
&=& f_{\rho}^T m_{\rho}^3 \frac{e^{(\lambda)}\cdot z }{2 (p \cdot z)}
    [ p_\alpha p_\mu g^\perp_{\beta\nu} 
     -p_\beta p_\mu g^\perp_{\alpha\nu} 
     -p_\alpha p_\nu g^\perp_{\beta\mu} 
     +p_\beta p_\nu g^\perp_{\alpha\mu} ] 
      {\cal T}(v,pz)
\nonumber\\
&+& f_{\rho}^T m_{\rho}^2
    [ p_\alpha e^{(\lambda)}_{\perp\mu}g^\perp_{\beta\nu}
     -p_\beta e^{(\lambda)}_{\perp\mu}g^\perp_{\alpha\nu}
     -p_\alpha e^{(\lambda)}_{\perp\nu}g^\perp_{\beta\mu}
     +p_\beta e^{(\lambda)}_{\perp\nu}g^\perp_{\alpha\mu} ]
      T_1^{(4)}(v,pz)
\nonumber\\
&+& f_{\rho}^T m_{\rho}^2
    [ p_\mu e^{(\lambda)}_{\perp\alpha}g^\perp_{\beta\nu}
     -p_\mu e^{(\lambda)}_{\perp\beta}g^\perp_{\alpha\nu}
     -p_\nu e^{(\lambda)}_{\perp\alpha}g^\perp_{\beta\mu}
     +p_\nu e^{(\lambda)}_{\perp\beta}g^\perp_{\alpha\mu} ]
      T_2^{(4)}(v,pz)
\nonumber\\
&+& \frac{f_{\rho}^T m_{\rho}^2}{pz}
    [ p_\alpha p_\mu e^{(\lambda)}_{\perp\beta}z_\nu
     -p_\beta p_\mu e^{(\lambda)}_{\perp\alpha}z_\nu
     -p_\alpha p_\nu e^{(\lambda)}_{\perp\beta}z_\mu
     +p_\beta p_\nu e^{(\lambda)}_{\perp\alpha}z_\mu ]
      T_3^{(4)}(v,pz)
\nonumber\\
&+& \frac{f_{\rho}^T m_{\rho}^2}{pz}
    [ p_\alpha p_\mu e^{(\lambda)}_{\perp\nu}z_\beta
     -p_\beta p_\mu e^{(\lambda)}_{\perp\nu}z_\alpha
     -p_\alpha p_\nu e^{(\lambda)}_{\perp\mu}z_\beta
     +p_\beta p_\nu e^{(\lambda)}_{\perp\mu}z_\alpha ]
      T_4^{(4)}(v,pz)
\nonumber\\
&+&\ldots
\label{eq:T3}
\end{eqnarray}
and 
\begin{eqnarray}
\langle 0|\bar u(z)
         gG_{\mu\nu}(vz) 
         d(-z)|\rho^-(P,\lambda)\rangle 
&=& i f_{\rho}^T m_{\rho}^2
 [e^{(\lambda)}_{\perp\mu}p_\nu-e^{(\lambda)}_{\perp\nu}p_\mu] S(v,pz),
\nonumber\\
\langle 0|\bar u(z)
         ig\widetilde G_{\mu\nu}(vz)\gamma_5 
         d(-z)|\rho^-(P,\lambda)\rangle
&=& i f_{\rho}^T m_{\rho}^2
 [e^{(\lambda)}_{\perp\mu}p_\nu-e^{(\lambda)}_{\perp\nu}p_\mu] 
  \widetilde S(v,pz).
\end{eqnarray}
Of these seven amplitudes, ${\cal T}$ is twist-3 and the other six 
are twist-4. 

The light-cone expansion of the non-local tensor operator can be written
to twist-4 accuracy as
\begin{eqnarray}
\lefteqn{\hspace*{-1.5cm}\langle 0|\bar u(x) \sigma_{\mu \nu} 
d(-x)|\rho^-(P,\lambda)\rangle =} \nonumber \\
&=& i f_{\rho}^{T} \left[ (e^{(\lambda)}_{\mu}P_\nu -
e^{(\lambda)}_{\nu}P_\mu )
\int_{0}^{1} \!du\, e^{i \xi P x}
\Bigg[\phi_{\perp}(u) +\frac{m_\rho^2x^2}{4} {\Bbb A}_T(u)\Bigg] \right. 
\nonumber \\
& &{}+ (P_\mu x_\nu - P_\nu x_\mu )
\frac{e^{(\lambda)} \cdot x}{(P x)^{2}}
m_{\rho}^{2} 
\int_{0}^{1} \!du\, e^{i \xi P x} {\Bbb B}_T (u) 
\nonumber \\
& & \left.{}+ \frac{1}{2}
(e^{(\lambda)}_{ \mu} x_\nu -e^{(\lambda)}_{ \nu} x_\mu) 
\frac{m_{\rho}^{2}}{P  x} 
\int_{0}^{1} \!du\, e^{i \xi P x} {\Bbb C}_T(u) \right],
\label{eq:OPE2}
\end{eqnarray}
where ${\Bbb B}_T$ and ${\Bbb C}_T$ are expressed in terms of the distribution 
amplitudes defined above as 
\begin{eqnarray}
   {\Bbb B}_T(u) &=& h_\parallel^{(t)}(u) -\frac{1}{2}\phi_\perp(u)-
              \frac{1}{2} h_3(u),
\nonumber\\
   {\Bbb C}_T(u) &=& h_3(u)-\phi_\perp(u),
\end{eqnarray}
and ${\Bbb A}_T$ can be  related to integrals of three-particle distribution 
functions using the equations of motion.

We introduce the notation, similar to Eq.~(\ref{eq:c12}):
\begin{eqnarray}
 {\Bbb B}_T^{(i)}(u) &=& -\int_0^u\!dv\,{\Bbb B}_T(v),
\nonumber\\
 {\Bbb C}_T^{(i)}(u) &=& -\int_0^u\!dv\,{\Bbb C}_T(v). 
\end{eqnarray}
\begin{table}
\renewcommand{\arraystretch}{1.4}
\addtolength{\arraycolsep}{3pt}
$$
\begin{array}{|c|cccc|}
\hline
V & \rho^\pm & K^*_{u,d} & \bar{K}^*_{u,d}& \phi\\ \hline
f_V [{\rm MeV}] & 198\pm 7  & 226 \pm 28& 226 \pm 28 & 254 \pm 3\\
f^T_V [{\rm MeV}] & 
\begin{array}{c} 160\pm 10 \\ 152\pm 9 \end{array}&
\begin{array}{c} 185\pm 10 \\ 175\pm 9 \end{array}&
\begin{array}{c} 185\pm 10 \\ 175\pm 9 \end{array}&
\begin{array}{c} 215\pm 15 \\ 204\pm 14 \end{array}
\\ \hline
a_1^\parallel & 0 & 
\begin{array}{c} 0.19 \pm 0.05 \\ 0.17\pm 0.04 \end{array}&
\begin{array}{c} -0.19 \pm 0.05 \\ -0.17\pm 0.04 \end{array}&
\phantom{-}0\\
a_2^\parallel & 
\begin{array}{c} 0.18 \pm 0.10 \\ 0.16\pm 0.09 \end{array}&
\begin{array}{c} 0.06 \pm 0.06 \\ 0.05\pm 0.05 \end{array}&
\begin{array}{c} \phantom{-}0.06 \pm 0.06 \\ 
                   \phantom{-}0.05\pm 0.05 \end{array}&
0\pm0.1\\
a_1^\perp & 0 & 
\begin{array}{c} 0.20 \pm 0.05 \\ 0.18\pm 0.05 \end{array}&
\begin{array}{c} -0.20 \pm 0.05 \\ -0.18\pm 0.05 \end{array}&
\phantom{-}0\\
a_2^\perp & 
\begin{array}{c} 0.20 \pm 0.10 \\ 0.17\pm 0.09 \end{array}&
\begin{array}{c} 0.04 \pm 0.04 \\ 0.03\pm 0.03 \end{array}&
\begin{array}{c} \phantom{-}0.04 \pm 0.04 \\ 
                  \phantom{-}0.03\pm 0.03 \end{array}&
 0\pm0.1\\ \hline
\delta_+ & 0 & 
\begin{array}{c} \phantom{-}0.24 \\ \phantom{-}0.22 \end{array}&
\begin{array}{c} 0.24 \\ 0.22 \end{array}&
\begin{array}{c} 0.46 \\ 0.41 \end{array}
\\
\delta_- & 0 & 
\begin{array}{c} -0.24 \\ -0.22 \end{array}&
\begin{array}{c} 0.24 \\ 0.22 \end{array}&
0 \\
\widetilde{\delta}_+ & 0 & 
\begin{array}{c} \phantom{-}0.16 \\ \phantom{-}0.13 \end{array}&
\begin{array}{c} 0.16 \\ 0.13 \end{array}&
\begin{array}{c} 0.33 \\ 0.27 \end{array}
\\
\widetilde{\delta}_- & 0 & 
\begin{array}{c} -0.16 \\ -0.13 \end{array}&
\begin{array}{c} 0.16 \\ 0.13 \end{array}&
0\\ \hline
\end{array}
$$
\caption[]{Masses and couplings of vector-meson distribution
  amplitudes, including SU(3)-breaking. In cases where two values 
are given, the upper one corresponds to the scale $\mu^2=1\,$GeV$^2$ 
and the lower one to $\mu^2 = m_B^2-m_b^2 = 4.8\,$GeV$^2$,
respectively.
We use $m_s(1\,{\rm GeV}) = 150\,$MeV and put the $u$ and $d$ quark
mass to zero.
}\label{tab:para}
$$
\begin{array}{|c|ccccccc|}\hline
& \zeta_3 & \omega_3^A & \omega_3^V & \omega_3^T & \zeta_4 & \zeta_4^T
& \tilde{\zeta_4^T}\\ \hline
V &
\begin{array}{c} 0.032\\ 0.023 \end{array}& 
\begin{array}{c} -2.1 \\  -1.8 \end{array}& 
\begin{array}{c} 3.8  \\   3.7 \end{array}& 
\begin{array}{c} 7.0  \\   7.5 \end{array}&
\begin{array}{c} 0.15 \\   0.13 \end{array}&  
\begin{array}{c} 0.10 \\   0.07 \end{array}&  
\begin{array}{c} -0.10\\  -0.07 \end{array}  
\\ \hline
\end{array}
$$
\caption[]{Couplings for twist-3 and 4 distribution amplitudes for
  which we do not include SU(3)-breaking. Renormalization scale as in
  the previous table.}\label{tab:para2}
\renewcommand{\arraystretch}{1}
\addtolength{\arraycolsep}{-3pt}
\end{table}
For the leading twist-2 distribution amplitude $\phi_\perp$ we use
\begin{equation}\label{eq:phiperp}
\phi_\perp(u) =  6 u\bar u \left[ 1 + 3 a_1^\perp\, \xi +
a_2^\perp\, \frac{3}{2} ( 5\xi^2  - 1 ) \right],
\end{equation}
with parameter values as specified in Table~\ref{tab:para}.
The expressions for higher-twist distributions given below correspond
to the simplest self-consistent approximation that satisfies all 
QCD equations of motion \cite{BBKT,BBS}:
\begin{itemize}
\item{} Three-particle distribution of twist-3:
\begin{equation}
{\cal T} (\underline{\alpha}) = 
540\, \zeta_3 \omega^T_3 (\alpha_d-\alpha_u)\alpha_d \alpha_u\alpha_g^2.
\end{equation}
\item{} Two-particle distributions of twist-3:
\begin{eqnarray}
h_\parallel^{(s)}(u) & = & 6u\bar u \left[ 1 + a_1^\perp \xi + \left( \frac{1}{4}a_2^\perp +
\frac{5}{8}\,\zeta_{3}\omega_3^T \right) (5\xi^2-1)\right]\nonumber\\
& & {}+ 3\, \delta_+\, (3 u \bar u + \bar u \ln \bar u + u \ln u) + 
3\,\delta_-\,  (\bar u
\ln \bar u - u \ln u),\label{eq:e}\\
h_\parallel^{(t)}(u) &= & 3\xi^2+ \frac{3}{2}\,a_1^\perp \,\xi (3 \xi^2-1)
+ \frac{3}{2} a_2^\perp\, \xi^2 \,(5\xi^2-3) 
+\frac{15}{16}\zeta_{3}\omega_3^T(3-30\xi^2+35\xi^4)\nonumber\\
& & {} + \frac{3}{2}\,\delta_+
\, (1 + \xi \, \ln \bar u/u) + \frac{3}{2}\,\delta_- \, \xi\, ( 2
+ \ln u + \ln\bar u )
\label{eq:hL}
\end{eqnarray}
\item{} Three-particle distributions of twist-4:
\begin{eqnarray}
 T^{(4)}_1(\underline{\alpha}) &=& T^{(4)}_3(\underline{\alpha}) ~~=~~0, 
 \nonumber\\
 T^{(4)}_2(\underline{\alpha}) &=& 
    30 \widetilde \zeta^T_{4}(\alpha_d-\alpha_u)\alpha_g^2,
 \nonumber\\
 T^{(4)}_4(\underline{\alpha}) &=& 
    - 30 \zeta^T_{4}(\alpha_d-\alpha_u)\alpha_g^2,
 \nonumber\\
 S(\underline{\alpha}) &=& 30 \zeta^T_{4}(1-\alpha_g)\alpha_g^2,
 \nonumber\\
 \widetilde S(\underline{\alpha}) &=& 
       30 \widetilde \zeta^T_{4}(1-\alpha_g)\alpha_g^2.
\end{eqnarray}
\item{} Two-particle distributions of twist-4:
\begin{eqnarray}
   h_3(u) &=& 6u(1-u)+5[\zeta^T_4+\widetilde \zeta^T_4](1-3\xi^2),
\nonumber\\
   {{\Bbb A}}_T(u) &=& 30 u^2(1-u)^2
   \Bigg[\frac{2}{5}+\frac{4}{3}\zeta^T_4-\frac{8}{3}
   \widetilde \zeta^T_4\Bigg].
\end{eqnarray}
\end{itemize}
The constants $\zeta^T_4$ and $\widetilde \zeta^T_4$
are defined as
\begin{eqnarray}
\langle 0|\bar u gG_{\mu \nu}d|\rho^-(P,\lambda)\rangle &=&
  if_\rho^T m_\rho^3 \zeta^T_4(
e^{(\lambda)}_{\mu}P_\nu - e^{(\lambda)}_{\nu}P_\mu),
\nonumber\\
\langle 0|\bar u g\widetilde G_{\mu \nu}i\gamma_5 
   d|\rho^-(P,\lambda)\rangle &=&
  if_\rho^T m_\rho^3 \widetilde \zeta^T_4(
e^{(\lambda)}_{\mu}P_\nu - e^{(\lambda)}_{\nu}P_\mu)
\end{eqnarray}
and have been estimated in \cite{BBK} from QCD sum-rules:
\begin{equation}
    \zeta^T_4 \simeq - \widetilde \zeta^T_4 \simeq 0.10.
\end{equation}
Other parameters are given in 
Table~\ref{tab:para}\footnote{In notations of Ref.~\cite{BBKT}
$\zeta_3^T \equiv (3/28)\zeta_3\omega_3^T$.}.
As in  the chiral-even case, we neglect SU(3)-breaking corrections
in twist-4 distributions.

\end{document}